\documentclass[12pt]{iopart} 

\begin{document}

\title{Dynamics of supercooled liquids: density fluctuations and Mode
  Coupling Theory} {\it $\!\!\!\!\!\!\!\!\!\!\!\!\!\!$ Preprint
  submitted to Journal of Physics: Condensed Matter (October 2001).}

\author{E. Zaccarelli$^1$, G. Foffi$^1$, P. De Gregorio$^1$,
F. Sciortino$^2$,  \\P. Tartaglia$^2$ and K. A. Dawson$^1$}
\address{$^1$ Irish Centre for Colloid Science and Biomaterials, Department of
Chemistry, University College Dublin, Belfield, Dublin 4, Ireland }
\address{$^2$  Dipartimento di Fisica, Universit\`{a} di Roma La Sapienza, 
Istituto Nazionale di Fisica della Materia, and INFM Center for
Statistical Mechanics and Complexity, Piazzale Aldo Moro 2, 00185
Roma, Italy}

\begin{abstract}
We write equations of motion for density variables that are equivalent
to Newtons equations.  We then propose a set of trial equations
parameterised by two unknown functions to describe the exact
equations.  These are chosen to best fit the exact Newtonian
equations. Following established ideas, we choose to separate these
trial functions into a set representing integrable motions of density
waves, and a set containing all effects of non-integrability. The
density waves are found to have the dispersion of sound waves, and
this ensures that the interactions between the independent waves are
minimised. Furthermore, it transpires that the static structure factor
is fixed by this minimum condition to be the solution of the
Yvon-Born-Green (YBG) equation. The residual interactions between
density waves are explicitly isolated in their Newtonian
representation and expanded by choosing the dominant objects in the
phase space of the system, that can be represented by a dissipative
term with memory and a random noise. This provides a mapping between
deterministic and stochastic dynamics.  Imposing the
Fluctuation-Dissipation Theorem (FDT) allows us to calculate the
memory kernel.  We write exactly the expression for it, following two
different routes, i.e. using explicitly Newtons equations, or instead,
their implicit form, that must be projected onto density pairs, as in
the development of the well-established Mode Coupling Theory (MCT).
We compare these two ways of proceeding, showing the necessity to
enforce a new equation of constraint for the two schemes to be
consistent.  Thus, while in the first `Newtonian' representation a
simple gaussian approximation for the random process leads easily to
the Mean Spherical Approximation (MSA) for the statics and to MCT for
the dynamics of the system, in the second case higher levels of
approximation are required to have a fully consistent theory.
\end{abstract}

\section{Introduction}  
The field of equilibrium statistical mechanics was founded on the
simple but remarkable proposition that motions generated by Newtonian
or Hamiltonian mechanics have very simple statistical properties if
there are sufficiently large numbers of interacting particles.
Specifically, the earliest developments of statistical mechanics
presupposed the so-called theorem of molecular chaos which assumed
that all parts of phase-space are equally probable
\cite{boltzmann}. This simple idea is at the basis of the ergodic
theorems, which state the equality of time and ensemble averages, the
latter having an explicit representation in the Boltzmann probability
distribution. Subsequent developments further strengthened the
foundations of statistical mechanics by weakening the molecular chaos
requirement, and the hypothesis of equal probability for all parts of
phase space, while retaining the formalism of statistical
mechanics. Other related developments allow to study the dynamics of
many particle systems using some sort of stochastic process that
mimics aspects of the chaotic motions of Newtonian mechanics. Indeed,
various advances have provided a firmer basis for relating Newtonian
mechanical averages to the statistical properties of stochastic
averages \cite{mori,hansen}.
  
Somewhat deeper problems arise when one wants to study those systems
where the parts of phase space explored over time are limited, and
limited in what is perhaps no longer a simple manner.  Already
approaching an equilibrium phase transition, a system becomes strongly
non-ergodic.  Indeed, when a phase transition occurs, by the breaking
of a symmetry for example, it is no longer appropriate to average over
all parts of the phase space in the probability averages of
Boltzmann. Instead we average only over those parts close to the
lowest, most probable energy state \cite{anderson}.  In such simple
cases the Boltzmann machinery, applied to a limited portion of phase
space, can still be applied.  However, when the ergodicity is lost in
the system in a less well defined manner it may not be possible to
find any straightforward generalisation of the Boltzmann distribution,
and one is forced to reconsider the possibility to perform time
averages with respect to another type of averaging.  The provision of
a well defined mechanism to do this is, in the arena of spin glasses,
one of the substantial achievements of that field
\cite{parisispinglass}.

Where dealing with particles, whether they be atoms, molecules, or
colloidal particles, we know at least that Newtonian dynamics, or some
equivalent stochastic process, remains faithful to the physical
picture, and averaging with respect to such a dynamics would be
satisfactory.  Examples of systems where these questions arise include
structural glasses \cite{angell}, the glass transition
\cite{debennature}, and many other cases where amorphous materials
such as gels \cite{fuchs,us,mechanical}, composites etc. are formed,
there being no prominent state to dominate the average, and render the
Boltzmann procedure acceptable. In such cases one of the few
approaches able to describe the dynamics of the system is the ideal
Mode Coupling Theory (MCT) \cite{bengt,gotze}. In fact, as we shall
see, one way of viewing MCT is as an approximate method of performing
Newtonian averages \cite{lettermct}. In this sense MCT is an important
exemplar of a new way of thinking about calculating statistical
dynamical averages, without depending on any form of ergodic
hypothesis, or on any assumption that the long time probability
distribution is the Boltzmann one.
  
In this paper, we shall present a formalism that connects Newtonian
averages taken with respect to initial positions and momenta to
stochastic averages taken with respect to noise distributions.  This
formalism has much in common with the Zwanzig-Mori formalism
\cite{mori}, but we here emphasise as much as possible the connection
and fidelity to Newtonian averages, and the possibility to explicitly
realise these averages within various approximations. 

An important aspect of the approach lies in its transparency. This
allows the possibility of an easy generalisation, for example, to
consider other types of collective variables, other than simply
density waves, as well as to represent non-integrable parts of motion
of the system not only by linear dissipation and random noise, as we
will discuss here.  We will also attempt, where possible, to probe the
physical meaning of the various steps, and above all the limitations
of the approximations made. One interesting consequence of our
approach is that ideal MCT emerges as a consistent and natural first
`mean-field' type approximation \cite{lettermct,thirumalai,crisanti}
to our theoretical framework, thereby providing an alternative route
to the MCT as a theory of dynamics, consistent with the Random Phase
Approximation (RPA) of statics. In this regard, we can now consider
MCT to be exact in the weak coupling limit.


Thus, we begin by writing equations of motion for the density
variables that are equivalent to Newtons equations. We then propose a
set of trial equations, parameterised by a set of unknown functions,
to describe these exact equations \cite{percus}.  These parameter
functions will have to be chosen to best fit the exact Newtonian
equations. 
We choose to separate the set of trial functions in the trial
equations of motion into a set representing integrable motions of
density waves (the dispersions), and a set containing all the
remaining effects of non-integrability. In so doing the density waves
are seen to have the dispersion of sound waves, and this choice
ensures that the interactions between the independent density waves is
minimised. Furthermore, it transpires that the structure factor, and
thereby the direct correlation function, is fixed by this minimum
condition to be the solution of the Yvon-Born-Green (YBG) equation
\cite{hansen,tosi}. As part of this process, residual interactions between  
the density waves are then explicitly isolated in their Newtonian
representation. We therefore have an explicit Newtonian representation
of the interaction force, and find a way of representing this force by
expanding the trial function.
        
Evidently it is possible to expand these interactions in a set of
functions that are best chosen to represent the dominant objects in
the phase space of the system that provide for loss of the pure
oscillatory motions of density waves.  The key question therefore is
the nature of the underlying phase space in the density representation
for that state of the system we seek to describe. It is emphasised
that these interaction parameters represent all the non-integrable
contributions of density waves, rather than particles. Thus, the phase
relations between density waves that are implied by the existence of
particles, as well as the interactions of these particles and the
resulting dissipation and other type of non-integrability of the
underlying Newtonian system must be accommodated by the trial
functions.  To make these ideas more concrete we will illustrate the
problem with a simple example of how the non-integrable parts of phase
space can be represented by a dissipation term with memory, as well as
a correlated noise to represent the more rapid fluctuations of the
density caused by effects like chaotic motions. This leads to a
generalised Langevin equation where the distribution of noise and the
memory kernel of the dissipative term are to be determined to reflect
the statistical properties of phase space.
   
It should be emphasised that, in choosing to approximate phase space
statistics partly with deterministic and partly with stochastic terms,
we will always naturally arrive at a stochastic process for the trial
density function equation. Therefore the idea of `best' representing
Newtonian phases space motion can only be implemented in some average
sense, rather than point-wise in particle position and momentum. This
will mean that we must relate, and equate, general averages over the
stochastic variables to those over the Newtonian space. The equality
of these two averages will, at each order, define the `goodness' of
the approximation.  These issues are discussed in section
\ref{sec:stochastic}.

Returning to the simple case where the trial forces or  
interactions are described by a noise and a memory kernel, to produce  
a conventional generalised Langevin equation it will be sufficient to  
apply the Fluctuation Dissipation Theorem of type two (FDT)  
\cite{kubo} to ensure that the system can reach equilibrium in the  
long time limit. It is natural to ask whether the use only of these
two terms in the trial forces is sufficient to represent Newtonian
dynamics, and we shall discuss the question at some length in section
\ref{subsec:mathematical}.  
  
We have earlier alluded to the matter of ensuring that averages over
density correlators are the same in Newtonian and stochastic
spaces. Since this involves a hierarchical process, there is the
practical need to truncate the program at some point.  One practical
possibility, outlined in section \ref{subsec:practical}, is to ensure
equality of the relevant correlators in the two spaces, and in
particular to ensure that the stochastic part of the force is
correlated properly with the density variables, again to some
appropriate order.  These considerations then lead us to a hierarchy
of equilibrium correlation function equations, the first level being
the YBG equation discussed above.
  
Following this, in section \ref{sec:exact}, we turn to the evaluation
of the exact memory kernel, using the two alternative representations
of the stochastic part of the force. This culminates in two exact
equations (\ref{eq:gamma}) and (\ref{eq:gamma2}), that represent the
central results of this paper.  
While these equations are exact, they are not closed. Therefore, in
sections \ref{sec:rep1}-\ref{sec:rep2} we discuss various aspects of
closure strategies. The most elegant closure found to date involves
assuming a Gaussian distribution for the noise, using the Newtonian
represntation, and this leads us (section \ref{sec:rep1}) directly to
the known ideal MCT for supercooled liquids.
  
It is important to note that, while our formulation and presentation
are different from earlier treatments, in many aspects there are
parallels and links to previous work.  Thus, in section \ref{sec:rep2}
we show how to deal with the implicit representation of the force, and
point out that only pairs of density variables are required for its
exact specification.  We also, in section \ref{subsec:rep2gaussian},
note that the Gaussian approximation can only be consistent up to
first order in density variables, and also that the stochastic force
cannot be consistently represented at the level of density pairs if it
is gaussian distributed.  Next, in section \ref{subsec:rep2singwi}, we
begin the process of seeking consistent approximations that permit the
representation of the stochastic force at the level of pairs of
density variables. As an example, we illustrate the
Singwi-Tosi-Land-Sjolander (STLS) \cite{stls} closure is introduced as
a step in this direction, but it also fails to be completely
satisfactory.

Finally in section \ref{sec:gotze}, by extracting the main elements
of the traditional derivation of MCT via projection operators, we
will attempt to clarify the assumptions made in our derivation of MCT
and in the traditional treatment \cite{gotze}.
  
  

\section{Density Equations of Motion}  
\noindent  
We begin with the definition of Newton's equations,   
\begin{equation}   
\ddot{{\bf r}}_j(t)=-  
\frac{1}{m}\frac{\partial {\cal V}}{\partial {\bf r}_j(t)}   
\label{newton}  
\end{equation}  
where ${\bf r}_j(t)$ are the positions of particles $j=1,N$ and $m$ is  
the particle mass, and we suppose that the total inter-particle  
potential ${\cal V}({\bf r}_1,...,{\bf r}_n)$ can be expressed as a sum of  
pair potentials,  
\begin{equation}   
{\cal V}=\frac{1}{2}\sum_{m,m'}{\it v}(|{\bf r}_m(t)-{\bf r}_{m'}(t)|)  
\label{pairpotential}   
\end{equation}  
The density of the $N$ particles located at positions ${\bf r}_j$ is
$\rho({\bf r},t)=\sum_{j=1}^N \delta({\bf r}-{\bf r}_j(t))$ and its
Fourier transform, which we name the density variable, is,
\begin{equation}   
\rho_{\bf k}(t)=\sum_{j=1}^N e^{i{\bf k}\cdot {\bf r}_j(t)}   
\label{rhok}  
\end{equation}  
and the structure factor of the system is defined as the density-density  
correlation,  
\begin{equation}   
S(k,t)=\frac{1}{N}\langle   
\rho_{-{\bf k}}(0)\rho_{{\bf k}}(t)\rangle 
\end{equation}  
which, for equal times, corresponds to the static structure factor
$S_k=\langle|\rho_{\bf k}|^2\rangle$.

\noindent  
We, thus, write the equations of motion corresponding to these
variables using the second time derivative of (\ref{rhok}),
\begin{equation}  
\ddot{\rho}_{\bf k}(t)=\sum_j i({\bf k}\cdot {\bf \ddot{r}}_j(t))   
e^{i{\bf k}\cdot {\bf r}_j(t)}-\sum_j ({\bf k}\cdot {\bf \dot{r}}_j(t))^2   
e^{i{\bf k}\cdot {\bf r}_j(t)}   
\label{der2rho}  
\end{equation}  
So far these equations are only formal equalities, but we can now  
eliminate the second time derivative of the positions using Newton's  
equations (\ref{newton}).  
To do this, we Fourier analyse equation (\ref{pairpotential}) to obtain,  
\begin{eqnarray}   
{\cal V}=\frac{1}{2V}\sum_{m,m'}\sum_{\bf k'}v_{k'}   
e^{-i{\bf k'}\cdot ({\bf r}_m(t)-{\bf r}_{m'}(t))} 
=\frac{1}{2V} \sum_{\bf k'}v_{k'}  
(\rho_{\bf k'}^*\rho_{\bf k'}-N\delta_{k',0})   
\label{fourierpot}
\end{eqnarray}  
where V is the total volume of the system and we have assumed that the
system is isotropic, and thus, the potential depends only on the
modulus of the wave-vector.

\noindent 
The derivative of the potential in Fourier space (\ref{fourierpot})
can be written as,
\begin{eqnarray}   
\!\!\!\!\!\!\!\!\!\!\!\!\!\!\!\!\!\!\!\!\!\!\!\!\!\!\!\!\!\!\!\!\!\!\!\!
\frac{\partial {\cal V}}{\partial r_j(t)}&=&  
\frac{1}{2V}\sum_{\bf k'}v_{k'}\left(\frac{\partial\rho_{\bf k'}^*}  
{\partial k'}\rho_{\bf k'}+  
\rho_{\bf k'}^* \frac{\partial\rho_{\bf k'}}{\partial k'}\right)=
\frac{1}{2V}\sum_m \sum_{\bf k'} v_{k'}  
(-i{\bf k'} e^{-i{\bf k'}\cdot ({\bf r}_j(t)-{\bf r}_m(t))}+c.c.)   
\end{eqnarray}  
so that the equations of motion for the density variables  
(\ref{der2rho}) now become,  
\begin{eqnarray}   \!\!\!\!\!\!\!\!\!\!  \!\!\!\!\!\!\!\!\!\!  
\ddot{\rho}_{\bf k}(t)&=&-\sum_j ({\bf k}\cdot   
{\bf \dot{r}}_j(t))^2 e^{i{\bf k}\cdot {\bf r}_j(t)}-
\frac{1}{mV}\sum_{j,m} \sum_{\bf k'}v_{k'} ({\bf k}\cdot {\bf k'})   
e^{-i{\bf k'}\cdot ({\bf r}_j(t)-{\bf r}_m(t))} e^{i{\bf k}\cdot {\bf  
r}_j(t)}\nonumber\\  \!\!\!\!\!\!\!\!\!\!\!\!\!\!\!\!\!\!\!\!  
&=&-\sum_j({\bf k}\cdot {\bf \dot{r}}_j(t))^2 e^{i{\bf k}\cdot   
{\bf r}_j(t)}-\frac{1}{mV}\sum_{\bf k'}v_{k'} ({\bf k}\cdot {\bf k'})   
\rho_{{\bf k}-{\bf k'}}(t)\rho_{\bf k'}(t) .
\label{der2rho2}  
\end{eqnarray}  
Note that equations (\ref{der2rho2}) constitute an exact set of
equations, that are equivalent to Newton's equations, but written in
terms of the density variables. Clearly the first term represents the
kinetic part of the motion, which is not simply expressable in terms
of density variables, while the second term, originating from the
interaction potential, indicates a coupling of the density pairs.

We note here that, in principle, to describe the underlying Newtonian
dynamics correctly, i.e. to include sufficient information in the
theory that would mimic positions and velocities of the $N$ particles,
one would need to consider many fields. In particular the most general
ones are usually considered the hydrodynamic variables of longitudinal
and transverse currents, along with a local temperature or entropy
variable \cite{landau}.  Until now, mode coupling type theories have
been based on longitudinal current, and therefore only the density
degree of freedom, and this is the approach we will also follow.
Extensions involving other fields are more complex, but, due to the
transparency of the approach, it should be quite straightforward at
least to write the correct equations of motion, as done in this and
the next sections for the density variables.

  
\section{Trial equations and Residual Force between Elementary Excitations}  
  
Equations (\ref{der2rho2}), being equivalent to Newton's equations,
are intractable as presently written. They must therefore be
approximated in some manner. To do this we consider the possibility to
isolate any simple degrees of freedom that can be easily treated,
especially if the real system can be considered as some sort of
perturbation around them.  Natural candidates for this treatment are
all integrable motions in the phase space. We will consider as
fundamental collective motions for describing the system and the
essential mechanisms by which its fluctuations relax the density
waves.



Thus, we propose to write equations (\ref{der2rho2}) in a simple `trial'  
form \cite{percus}.  Indeed, we consider the equations of motion, for  
density waves that have some interaction force between them, to be of  
the form,  
\begin{equation}   
\ddot{\rho}_{\bf k}(t)+\hat{\Omega}_{\bf k}\rho_{\bf k}(t)=  
\hat{\cal F}_{\bf k}(t)   
\label{trialeq}  
\end{equation}  
Here, we have isolated the linear term in $\rho_{\bf k}(t)$, that
represents the elementary excitations of the system, and we have
indicated by $\hat{\cal F}_{\bf k}(t)$ the interaction force between
them, this being a functional of the frequencies $\hat{\Omega}_{\bf
k}$ that have to be determined, as well as any other parameters to
represent the forces.  These elementary excitations can be interpreted
as a sort of `phonons'. Of course, these would be oscillating
perfectly with the frequencies $\hat{\Omega}_{\bf k}$ if there was no
interaction force between them.  Thus, they represent integrable
motions and we also assume that, for the purpose of describing a
liquid, these are the only collective motions that need to be taken
into account.
The residual interaction force between these `phonons', $\hat{\cal  
F}_{\bf k}(t)$, is the responsible for the deviation of the density waves  
from a perfect oscillatory behaviour. The smaller it is, the simpler  
the solutions $\rho_{\bf k}(t)$ of equations (\ref{trialeq}) will be.  
Generally, in a liquid, this interaction is not negligible,   
as it would be in a crystal.   
  
Now, we apply a variational principle to minimise this interaction  
\cite{percus,zwanzig}. Our variational parameters are the dispersions    
$\hat{\Omega}_{\bf k}$. 
We require,  
\begin{equation}   
\frac{\partial \langle|\hat{\cal F}_{\bf k}|^2  
\rangle_{\scriptscriptstyle \cal N}}  
{\partial \hat{\Omega}_{\bf k}}=0   
\label{minimo}   
\end{equation}  
and thereby calculate the optimal dispersion $\Omega_{\bf k}$, that  
satisfies this condition  and thus results in the least  
residual interaction between density waves.  \\
Here the averages are performed over initial conditions of positions  
and velocities, $\{r_n(0),\dot{r}_n(0)\}$, or equivalently of density  
variables and their velocities, $\{\rho_{\bf k}(0),\dot{\rho}_{\bf  
k}(0)\}$, via the definition (\ref{rhok}). Therefore we have indicated  
them with the notation $\langle \cdot \cdot \cdot  
\rangle_{\scriptscriptstyle \cal N}$ to denote that they are Newtonian  
averages.  We have,  
\begin{eqnarray}   
\!\!\!\!\!\!\!\!\!\!\!\!\!\!\!\!\!\!\!\!\!\!\!\!\!\!\!\!\!\!\!\!\!\!
\frac{\partial \langle|\hat{\cal F}_{\bf k}|^2  
\rangle_{\scriptscriptstyle \cal N}}  
{\partial \hat{\Omega}_{\bf k}}\!&=& \! 
\frac{\partial \langle|\ddot{\rho}_{\bf k}(t)+  
\hat{\Omega}_{\bf k}\rho_{\bf k}(t)|^2\rangle_{\scriptscriptstyle \cal N}}  
{\partial \hat{\Omega}_{\bf k}}\!=
\!\langle\rho_{-{\bf k}}(\ddot{\rho}_{\bf k}+ 
\hat{\Omega}_{\bf k}\rho_{\bf k})\rangle_{\scriptscriptstyle \cal N}\!+\!  
\langle(\ddot{\rho}_{-{\bf k}}+\hat{\Omega}_{\bf k}  
\rho_{-{\bf k}})\rho_{\bf k}\rangle_{\scriptscriptstyle \cal N}\!=\!0   
\label{minimize}  
\end{eqnarray}  
so that the solution is,  
\begin{equation}   
\Omega_{\bf k}=\frac{\langle|\dot{\rho}_{\bf k}|^2  
\rangle_{\scriptscriptstyle \cal N}}  
{\langle|{\rho}_{\bf k}|^2\rangle_{\scriptscriptstyle \cal N}}   
\label{dispersion1}  
\end{equation}  
This is a general result. Indeed, it was shown by Zwanzig  
\cite{zwanzig} to be valid for the frequencies associated with any  
elementary excitations in a fluid, in our case $\rho_{\bf k}(t)$.  By  
using the definitions of density variables (\ref{rhok}), one easily obtains, 
\begin{equation}  
\Omega_{\bf k}=\frac{k^2}{\beta m S_k}  
\label{dispersion}  
\end{equation}  
The same result (\ref{dispersion}) can also be easily found by  
evaluating the short-time limit of equations (\ref{der2rho2}), after  
multiplying them for ${\rho}_{-{\bf k}}(0)$ and performing the  
thermodynamical average \cite{hansen,gotze}.  The particular  
derivation we have presented here is instructive in showing how the  
collective motions are being removed from further consideration in  
the dynamics of the system, as all of the attention from now on can be  
focused on the residual interactions between them.  
We can now calculate this explicitly.  By adding to both sides of
equations (\ref{trialeq}) the term $\Omega_k\rho_{\bf k}(t)$, we have
an explicit form for the minimum residual force ${\cal F}_{\bf
k}(t)=\min{\hat{\cal F}}_{\bf k}(t) $. Thus,
\begin{eqnarray}   
\!\!\!\!\!\!\!\!\!\!\!\!\!\!\!\!\!\!\!
{\cal F}_{\bf k}(t)=\Omega_k\rho_{\bf k}(t)-  
\sum_j ({\bf k}\cdot {\bf \dot{r}}_j(t))^2 e^{i{\bf k}\cdot {\bf r}_j(t)}  
-\frac{1}{mV}\sum_{\bf k'}v_{k'} ({\bf k}\cdot {\bf k'})   
\rho_{{\bf k}-{\bf k'}}(t)\rho_{\bf k'}(t) .
\label{bareforce}  
\end{eqnarray}  
Clearly at some point we will have to represent this force in a  
simpler manner, if we are to make progress.  
  

\section{The Yvon-Born-Green Equation}  
\noindent  
It is of interest to note the fact that the best excitations (in the  
sense of minimising residual interactions between them) also implies  
some conditions on the interaction force.  Indeed, if we refer back to  
the minimisation of equation (\ref{minimize}), we can see that there  
an orthogonality condition is implied, i.e.  
\begin{equation}   
\langle \rho_{-{\bf k}}(t){\cal F}_{\bf k}(t)  
\rangle_{\scriptscriptstyle \cal N}=0   
\label{orthog}  
\end{equation}  
This is the statement that the instantaneous force at every time must
be orthogonal to the density waves themselves. 
This is an exact condition that should be enforced in
any type of `trial' equations of motion developed to describe Newton's
equations.  Thus, it must also be applied to our equations, with
minimised residual force,
\begin{equation}   
\ddot{\rho}_{\bf k}(t)+\Omega_{\bf k}\rho_{\bf k}(t)=  
{\cal F}_{\bf k}(t)   
\label{trialeqmin}  
\end{equation}  
Of course, this equation can be used to determine the equilibrium
correlation functions. Indeed, using the explicit expression of the
residual force (\ref{bareforce}), the orthogonality condition may be
written out explicitly as,
\begin{eqnarray}   \!\!\!\!\!\!\!\!\!\!\!\!\!\!
\!\!\!\!\!\!\!\!\!\!\!\!\!\!\!\!\!\!\!
\Omega_{\bf k}\langle|\rho_{\bf k}|^2
\rangle_{\scriptscriptstyle \cal N}\!-\!   
\langle\sum_j({\bf k}\cdot {\bf \dot{r}}_j(t))^2 
e^{i{\bf k}\cdot {\bf r}_j(t)}\rangle_{\scriptscriptstyle \cal N}\!-\! 
\frac{1}{mV}\sum_{\bf k'}v_{\bf k'} ({\bf k}\cdot {\bf k'}) \langle  
\rho_{-{\bf k}}(t) \rho_{{\bf k}-{\bf k'}}(t)\rho_{\bf k'}(t)  
\rangle_{\scriptscriptstyle \cal N}\!=\!0 \end{eqnarray}   
Now we note that the averages involved here are equilibrium averages,  
and as such they involve nothing more than the Boltzmann distribution.  
Thus, the averages of positions and momenta (or velocities) factor and  
can be carried out separately to yield,  
\begin{eqnarray}   
-\frac{k^2 }{\beta m } \left(1-\frac{1}{S_k}\right) N S_k= 
\frac{1}{mV}\sum_{\bf k'}v_{\bf k'} ({\bf k}\cdot {\bf k'})   
\langle \rho_{-{\bf k}}(t) \rho_{{\bf k}-{\bf k'}}(t)\rho_{\bf k'}(t)  
\rangle_{\scriptscriptstyle \cal N}   
\end{eqnarray}  
and defining the direct correlation function, $c_k=
\frac{1}{n}\left(1-\frac{1}{S_k}\right)$, where $n=N/V$ is the number
density of the system, we find,
\begin{equation}   
c_ k= -\beta v_k-
\frac{\beta}{k^2 N^2 S_k}  
\sum_{{\bf k'}\neq {\bf k}} v_{\bf k'} ({\bf k}\cdot {\bf k'})   
\langle \rho_{-{\bf k}}(t) \rho_{{\bf k}-{\bf k'}}(t)\rho_{\bf k'}(t)  
\rangle_{\scriptscriptstyle \cal N}   
\label{ybg}  
\end{equation}  
where we have extracted from the sum the term corresponding to ${\bf
k}={\bf k'}$ since it corresponds to simply $-\beta v_k$. Condition
(\ref{ybg}) constitutes a set of equations for the equilibrium fluid.
It can also be shown \cite{tosi} that Eq.\ref{ybg} is simply the
representation in Fourier space of the YBG equation.
Thus, equation (\ref{ybg}) gives a relation for the pair (static
structure factor) and triplet density correlation functions in terms
of the potential of the system. For example, if one could determine
the static structure factor by ensuring this condition, one would have
also determined the set of frequencies $\Omega_k$ that appear in the
trial equations (\ref{trialeqmin}).
  

\section{Transition from a Deterministic   
to a Stochastic Dynamics}   
\label{sec:stochastic}  
\noindent  
As we commented under equation (\ref{bareforce}), it is necessary to
approximate in some way the residual force to make further
progress. Broadly speaking, it has always been assumed that such forces
can be separated into a slowly varying and a rapidly varying part
\cite{mori,hansen,oppenheim}. If we decide to represent the rapidly
varying part as some type of noise, then the trial equations of motion
for the density variables will become stochastic equations.

In this section we will examine the transition from a deterministic to
a stochastic dynamics in its various aspects. Firstly, in subsection
\ref{subsec:physical}, we will propose a simple form for the residual
minimised interactions ${\cal F}_{\bf k}(t)$ between density waves,
basing ourselves on standard mathematical formalisms \cite{mori}. This
will allow us to write the density equations of motion
(\ref{der2rho2}) in the form of a set of generalised Langevin
equations. Then, in subsection \ref{subsec:mathematical} we will focus
on the mathematical constraints that must be imposed to our stochastic
process to faithfully reproduce its `true' Newtonian dynamics. 
Finally, in subsection \ref{subsec:practical} we propose a practical
way of implementing constraints between the Newtonian and the
stochastic space.

  
\subsection{The Physical Picture}  
\label{subsec:physical}  
So far, we have written a set of trial equations (\ref{trialeqmin}),
where all of the collective parts of the motion have been isolated
and, at the same time, the interactions or effective forces between
these collective motions have been minimised. The outcome is a set of
collective motion frequencies (\ref{dispersion}) that are related to
the structure factor, and that this structure factor is the solution
of the YBG equation.  \\
Now, the effective interactions, or trial forces, in equation
(\ref{bareforce}), must be approximated to accommodate the residual
physical interactions, but preserving the mathematical constraints,
such as (\ref{orthog}) discussed above, between them.  A central
observation is that, in the absence of the coupling terms ${\cal
F}_{\bf k}(t)$, the collective motions would have infinite
lifetimes. On the contrary, in a real system, depending on its
details, we have different types of effects, tending either to cause
the collective phonon-type motions to decay or to make them more long
lived. Indeed, the same system, for different temperatures or
densities, may exhibit such different tendencies, so that it will be
important to design trial forces that are flexible enough to
accommodate the particular phenomena of interest, as for example, the
glass transition.
   
Let us first deal with the issue of particle versus wave motion. Of
course, if we are able to solve Newton's equations exactly in either
their familiar particle or their density representation there will be
no difference. That is, we may represent all of the phenomena equally
well in terms of particles, or density waves, providing every step is
exact. Inevitably approximations will be introduced and then the
representation that we choose will be important. For smaller
wave-vectors we know that a substantial part of the motion is
collective. For larger wave-vectors we expect that aspects of the
independent and uncoordinated nature of the single-particle motion
will be dominant.  It should be noted that at long wavelengths we are
dealing with waves where many particle collisions occur and where the
longer density waves are locally in equilibrium.  This is also
reflected in the fact that we seek to model the collective
oscillations using frequencies that involve the equilibrium structure
factor. Therefore it is only part of the collisional spectrum that
contributes to the decay of the density variables, the remainder
having contributed to local equilibration of the density waves.
  
We now turn to the question of how the remaining effects of finite
lifetimes of these waves can be described. Also, we recall that it
will be important to retain at least some information on the relative
phases of the density waves, so as to have an adequate description of
particle, rather than wave, properties.
  
We assume, as usual, that the finite lifetime arises from correlated
phenomena, as well as more chaotic effects. It is thus conventional to
separate the residual interactions into two types of contribution,
following the ideas of the Zwanzig-Mori formalism \cite{mori}. The
first term involves a first derivative in density, so it will clearly
cause a decay of the oscillating density waves. However, by using a
memory kernel, we can parameterise some aspects of the history of the
system. The second term is a general random noise that originates a
more `chaotic' behaviour for the system.  Of course, such a separation
is not a final and complete one, but it is conventional, and we will
accept it for the current purposes.

By this statement we can represent our residual force
(\ref{bareforce}) in the most general form using these contributions
to non-integrability,
\begin{equation}   
{\cal F}_{\bf k}(t)=-\int_0^t \gamma_{\bf k}(t-t')   
\dot{\rho}_{\bf k}(t') dt' + \eta_{\bf k}(t)   
\label{expansion}  
\end{equation}  
where $\gamma_{\bf k}(t)$ is the memory function of the system and  
$\eta_{\bf k}(t)$ is the noise.  
The form (\ref{expansion}) for the residual force must, of course,  
satisfy the orthogonality condition (\ref{orthog}). 
  
We now return briefly to the effects of having a memory kernel in
equation (\ref{expansion}). Clearly, by selecting various forms for
$\gamma_{\bf k}(t)$ we are able to control the means by which
dissipation is expressed in the system. Thus, short-time effects that
would, for a normal liquid, lead to more rapid decay of the waves, can
be incorporated. Also, more extensive collective effects involving
many particles can be incorporated, in an averaged manner. Finally,
again in some averaged manner, we can incorporate those aspects of
phase coherences of different density waves, that are necessary to
retain a minimal picture of particle motions. Indeed, potentially we
are using the memory kernel to accommodate a number of different
physical effects.
  
By this expansion, we have made a transition from a deterministic  
description of the system, based on equations (\ref{trialeqmin}) and  
(\ref{bareforce}), to a stochastic one, where the full equations of  
motions are expressed through the generalised Langevin equations,  
\begin{equation}   
\ddot{\rho}_{\bf k}(t)+\Omega_k \rho_{\bf k}(t)+  
\int_0^t \gamma_{\bf k}(t-t') \dot{\rho}_{\bf  
k}(t') dt' = \eta_{\bf k}(t)   
\label{langevin}  
\end{equation}  
with $\Omega_{\bf k}$ given by (\ref{dispersion})and $\gamma_{\bf
k}(t)$ still to be determined.  Introducing the normalised density
correlator, defined as $\Phi_k(t)=S_k(t)/S_k$, the exact equations of
motion clearly are,
\begin{equation}   
\ddot{\Phi}_{\bf k}(t)+\Omega_{\bf k}\Phi_{\bf k}(t)+  
\int_0^t \gamma_{\bf k}(t-t') \dot{\Phi}_{\bf k}(t') dt' = 0  
\label{correlatoreqns}  
\end{equation}  
where the causality relation $ \langle \rho_{-{\bf k}}(0)\eta_{\bf
k}(t) \rangle=0$ for $t>0$ must hold.

In our system the fluctuations, represented by $\eta_{\bf k}(t)$, and the  
dissipations, expressed by $\gamma_{\bf k}(t)$, are not independent,  
and, in particular, they must satisfy the fluctuation-dissipation  
theorem (FDT), or else no Boltzmann equilibrium distribution would be  
reached by the Langevin equation implied by equations  
(\ref{langevin}). Thus, we have,  
\begin{equation}   
\frac{\langle \eta_{-{\bf k}}(t)\eta_{\bf k}(t')\rangle}  
{\langle|\dot{\rho}_{\bf k}|^2\rangle}= \gamma_{\bf k}(t-t')   
\label{fdt}  
\end{equation}  
This relation fixes the memory function. It can be easily shown that
this relation is directly implied by our generalised Langevin
equations (\ref{langevin}), with the only assumption of
equilibrium. To describe out-of-equilibrium dynamics, a generalization
of this formalism, and consequently of FDT, is necessary, as we
discussed in \cite{aging}.

We now want to clarify the question of time-translational invariance
of our stochastic process (\ref{langevin}).  Indeed, there is an
explicit dependence on the arbitrary initial time $t=0$, at which
memory effects start to be taken into account, due to the crucial
choice for the residual force (\ref{expansion}).  This is done for
obvious physical reasons.
If we consider another Langevin process of the form,
\begin{equation} 
\ddot{\rho}_{\bf k}+\Omega_k \rho_{\bf k}(t)+ 
\int_{-\infty}^t \gamma_{\bf k}(t-t') \dot{\rho}_{\bf 
k}(t') dt' = \tilde{\eta}_{\bf k}(t) 
\label{langevintti} 
\end{equation} 
which is now clearly time-translational invariant, we have the
following relation between our true random noise $\eta_{\bf k}(t)$ of
equations (\ref{langevin}), and this new, auxiliary, noise
$\tilde{\eta}_{\bf k}(t)$ \cite{kubo},
\begin{equation} 
\eta_{\bf k}(t)=\tilde{\eta}_{\bf k}(t)-\int_{-\infty}^0 dt' 
\gamma_{\bf k}(t-t') 
\dot{\rho}_{\bf k}(t') 
\label{randomtti} 
\end{equation} 
it is sufficient to ask our non-time-translational invariant
noise $\eta_{\bf k}(t)$ to obey the initial conditions,
\begin{eqnarray} 
&&\langle \rho_{-{\bf k}}(0)\eta_{\bf k}(t)\rangle=0 \ \ \ \ t\geq0\nonumber\\ 
&&\langle \dot{\rho}_{-{\bf k}}(0)\eta_{\bf k}(t)\rangle=0 \ \ \ \ t>0 
\end{eqnarray} 
as well as that its autocorrelation function is the same as the one
for $\tilde{\eta}_{\bf k}(t)$. Of course, we require the memory kernel,
\begin{equation} 
\langle \eta_{-{\bf k}}(0)\eta_{\bf k}(t)\rangle=
\langle\tilde{\eta}_{-{\bf k}}(t_1) 
\tilde{\eta}_{\bf k}(t_1+t)\rangle =\gamma_{\bf k}(t)
\end{equation} 
to ensure that the two stochastic processes (\ref{langevin}) and
(\ref{langevintti}) are completely equivalent \cite{kubo}. Thus,
causality is also correctly built in our equations (\ref{langevin}).
 
This means that, though the random noise $\eta_{\bf k}(t)$ itself is
not time-translational invariant, the whole process, described by
equations (\ref{langevin}), is. In particular, we have that the
residual force ${\cal F}_{\bf k}(t)$ (\ref{expansion}), which is the
sum of the random noise plus the conventional memory term, starting
from $t=0$, does not depend on the initial time.  Indeed, it is
important to clarify this aspect, as the orthogonality condition
(\ref{orthog}) between the density and the force ${\cal F}$ at equal
times, implies,
\begin{equation} 
\langle \rho_{-{\bf k}}(t){\cal F}_{\bf k}(t)\rangle= 
\langle\rho_{-{\bf k}}(0){\cal F}_{\bf k}(0)\rangle=\langle 
\rho_{-{\bf k}}(0)\eta_{\bf k}(0)\rangle=0 
\end{equation} 
since at the initial time, we have ${\cal F}_{\bf k}(0)=\eta_{\bf
k}(0)$. But we cannot simply say that also $\langle \rho_{-{\bf
k}}(t)\eta_{\bf k}(t)\rangle$ is zero, because of the non-validity of
the time-translational invariance for the noise $\eta_{\bf k}(t)$. In
fact, for any time $t$, the orthogonality condition just implies,
\begin{equation} 
\langle \rho_{-{\bf k}}(t)\eta_{\bf k}(t)\rangle= 
\int_0^t\gamma_{\bf k}(t-t')\langle\rho_{-{\bf k}}(t) 
\dot{\rho}_{\bf k}(t')\rangle 
\label{orthog2} 
\end{equation} 
and, in particular, each of these terms is zero in the trivial case of
a $\delta$-function memory kernel, for which, of course, the equations
motion become local in time.

  
\subsection{The Mathematical Constraints}   
\label{subsec:mathematical}  
\noindent   
In this section we will discuss the mathematical definition of our
stochastic process, and the implied constraints. This will shed some
light into how good our Langevin process (\ref{langevin}) is in
reproducing the Newtonian process in equations (\ref{der2rho2}).
Thus, the stochastic process $\eta_{\bf k}(t)$ can be defined via the  
generating function,  
\begin{eqnarray}  \!\!\!\!\!\!
\langle e^{i \sum_j\! u_{{\bf k}_j}(t_j)\eta_{{\bf k}_j}(t_j)}  
\rangle_{ \{\eta_{\bf k}\}}=
\langle e^{i \sum_j \!\!\!u_{{\bf k}_j}(t_j)[{\cal F}_{{\bf k}_j}(t_j)+  
\int_0^{t_j}\!\! \gamma_{{\bf k}_j}(t_j-t')\dot{\rho}_{{\bf k}_j}(t') dt']}   
\rangle_{ \{ {\bf r}_n(0), \dot{\bf r}_n(0) \} }  
\label{generating}  
\end{eqnarray}  
where the average on the lhs corresponds now to an average
over the noise probability distribution. Equation (\ref{generating})
is, in principle, a full definition of the random process, because a
stochastic process is completely defined by all its moments. Indeed,
every correlation function of the noise, considered as some moment of
the distribution, could be calculated by deriving it with respect to
the auxiliary parameters $u_{\bf k}$ and it is thereby related
directly to an average with respect to Newton's equations. These
relations are only just formal, but they do place a clear set of
constraints on the stochastic process that we choose to mimic Netwonian
dynamics.
  
Besides this relation between the stochastic process and Newton's
equations, there is also, via equation (\ref{langevin}), a mapping
between the density variables of the Langevin process and the noise.
Indeed, Laplace transforming equations (\ref{langevin}) and solving
with respect to the density, we have,
\begin{equation}   
\tilde{\rho}_{\bf k}(z)=\frac{[-iz+\tilde{\gamma}_{\bf k}(z)]\rho_{\bf k}(0)+  
\dot{\rho}_{\bf k}(0)+\tilde{\eta}_{\bf k}(z)}  
{-z^2-iz\tilde{\gamma}_{\bf k}(z)+\Omega_k}   
\label{laplacequations}  
\end{equation}  
where $\rho_{\bf k}(0)$ and $\dot{\rho}_{\bf k}(0)$ are the initial  
conditions in time of our variables, and we have used the definition  
of Laplace transform,  
\begin{equation}   
\tilde{\rho}_{\bf k}(z)=\int_0^{\infty}dt e^{izt}\rho_{\bf k}(t)   
\label{laplace}  
\end{equation}  
$z$ being a complex number.  Thus, now averages over  
density variables in the deterministic language can be replaced by  
averages over the noise $\eta_{\bf k}(t)$ in the stochastic one, or their  
equivalent density averages.  \\
Equation (\ref{fdt}) defines the autocorrelation function of the noise
distribution, but to give a complete definition of the stochastic
process we also need to define the mean value of the distribution. To
do so, we can use the mapping with density variables, which implies the
further condition $\langle \eta_{\bf k}(t)\rangle=0$.

Now, in principle, to create the noise distribution we should use the
definition of the stochastic process, via the generating function in
(\ref{generating}), and by taking derivatives of this, calculate the
noise distribution moments satisfying this equality up to any desired
accuracy. From these constraints, then, one can determine the density
correlators.  The level at which one truncates this program, via some
closure, will be the level of approximation of the theory.  Evidently,
if one followed this program, all averages over any function of
density variables taken in the Newtonian space could be explicitly
written in terms of averages over density variables in the stochastic
averages.

However, in practice we are not interested in the correlators of
arbitrarily high order, nor are we interested in arbitrary functions
of density variables.  Indeed, what we still need to determine to have
a complete theory is the memory kernel $\gamma_{\bf k}(t)$, which must
satisfy the FDT relation (\ref{fdt}). We will illustrate this in the
following of the chapter. Nonetheless, for the present purpose, it is
interesting to anticipate that the result for the memory will depend
only on 2-, 3-, and 4-point density correlators for unequal times, and
thus these are the only averages that need to be equal in the two
spaces, Newtonian and stochastic respectively. Therefore, we should
focus most attention only on these correlators.

We can write explicitly the implied equalities for 2- and 4-point
correlation functions at unequal times for the noise, derived from
(\ref{generating}), i.e.
\begin{equation}
\!\!\!\!\!\!\!\!\!\!\!\!\!\!\!\!\!\!\!\!\!\!
\langle \eta_{-{\bf k}_1}(0)\eta_{{\bf k}_2}(t) 
\rangle_{ \{\eta_{\bf k}\}}= \left\langle {\cal F}_{-{\bf k}_1}(0) 
\left[ {\cal F}_{{\bf k}_2}(t)+\int_0^{t} dt' 
\gamma_{{\bf k}_2}(t-t')\dot{\rho}_{{\bf k}_2}(t')\right] \right
\rangle_{ \{ {\bf r}_n(0), \dot{\bf r}_n(0) \} } \label{ff}
\end{equation}
\begin{eqnarray}
\!\!\!\!\!\!\!\!\!\!\!\!\!\!\!\!\!\!\!\!\!\!\!\!\!\!\!\!\!\!\!\!
&&\langle \eta_{-{\bf k}_1}(0)\eta_{-{\bf k}_2}(0)\eta_{{\bf k}_3}(t)
\eta_{{\bf k}_4}(t)\rangle_{ \{\eta_{\bf k}\}}= \left\langle {\cal
F}_{-{\bf k}_1}(0){\cal F}_{-{\bf k}_2}(0) \left[ {\cal F}_{{\bf
k}_3}(t)+\int_0^{t} dt' \gamma_{{\bf k}_3}(t-t')\dot{\rho}_{{\bf
k}_3}(t')\right] \right.
\nonumber\\ 
\!\!\!\!\!\!\!\!\!\!\!\!\!\!\!\!\!\!\!\!\!\!\!\!\!\!\!\!\!\!\!\!
&& \ \ \ \
\ \ \ \ \ \ \ \ \ \ \ \ \ \ \ \ \ \ \ \ \ \ \ \ \ \ \ \ \ \ \ \ \ \ \
\ \ \ \ \ \ \ \left.  
\left[ {\cal F}_{{\bf k}_4}(t)+\int_0^{t} dt'
\gamma_{{\bf k}_4}(t-t')\dot{\rho}_{{\bf k}_4}(t')\right]
\right\rangle_{ \{ {\bf r}_n(0), \dot{\bf r}_n(0) \} } 
\label{ffff}
\end{eqnarray} 
These equations define the stochastic averages of the correlations, or
what is equivalent, the definition of the stochastic process in terms
of the Newtonian dynamics, to the appropriate order. But we can
further note that, using the definition of ${\cal F}_{\bf k}(t)$ in
terms of the density variables (\ref{bareforce}) and substituting into
this the expression for the densities in terms of noise
(\ref{laplacequations}), we obtain a hierarchy of constraints for
the noise distributions. Indeed, without writing explicitly all of
these contraints, it is easy to see that equation (\ref{ff}) implies a
relationship for $\langle \eta\eta \rangle$ in terms of
$\langle\eta\eta\eta \rangle$ and $\langle\eta\eta\eta\eta
\rangle$. Similarly, equation (\ref{ffff}) implies relations between
$\langle \eta\eta \eta\eta\rangle$ and higher order correlations of
the noise up to order $\langle \eta\eta\eta\eta\eta\eta
\eta\eta\rangle$.

\noindent
These constraints should be enforced for any implementation of
Mori-type formalisms, with respect to the appropriate Newtonian
dynamics.  Then, practically, all closures of the theory involve
violation of these constraints at some order of noise moments.

\subsection{Equivalent Representations of the Random Force and 
Practical Implementation of Constraints}   
\label{subsec:practical}  
\noindent   
We have seen so far that the right constraints to implement for the
stochastic process to represent faithfully the Newtonian process
follow directly from the definition of the random noise
(\ref{generating}). More practically, we have written explicitly the
type of constraints that will be requested for our simple theory,
based on the expansion (\ref{expansion}) for the residual interactions
${\cal F}_{\bf k}(t)$, in (\ref{ff}) and (\ref{ffff}).  Now, if one
correctly implemented the constraints of type (\ref{ff}) and
(\ref{ffff}), as well the similar ones for the triplet correlation
averages that will appear in the expression for the memory kernel in
the next section, the transition from Newtonian to stochastic averages
would be successfully achieved for our formalism.  Thus, in
particular, one could consider the two sides of the exact density
equations (\ref{der2rho2}) to be equal when averaged with respect to
any order of the density variables using the Newtonian phase space, or
via the relation (\ref{laplacequations}), to any order of the noise
variables.

However, the execution of this program, i.e. the exact implementation
of (\ref{ff}) and (\ref{ffff}), is generally difficult to pursue, and
it may be of interest to ensure less extensive conditions, but ones
which address the most important physical effects. The fact that the
two sides of equations (\ref{der2rho2}) should be equal with respect
to the averaging process, to some chosen order of the density
variables, at least ensures that the underlying Newtonian dynamics is
reasonably reflected in the stochastic process. Thus, considering
equation (\ref{expansion}), it is clear that the random force
$\eta_{\bf k}(t)$ can be explicitly written as,
\begin{equation}   
\eta_{\bf k}(t)={\cal F}_{\bf k}(t)+\int_0^t \gamma_{\bf k}(t-t')   
\dot{\rho}_{\bf k}(t') dt'    
\label{randomf}  
\end{equation}  
but  the residual force can be expressed either in its explicit form, given  
in (\ref{bareforce}), or in its implicit one, given by  
(\ref{trialeqmin}). Thus, we have the two equivalent representations  
for the noise,  
\begin{eqnarray}  
\!\!\!\!\!\!\!\!\!\!\!\!\!\!\!\!\!\!\!\!\!\!\!\!\!\!\!\!\!\!\!\!\!\!\!\!
\!\!\!\!\eta^{(I)}_{\bf k}(t)\!&=&\!\Omega_k\rho_{\bf k}(t)\!-\! \! 
\!\sum_j ({\bf k}\!\cdot\! {\bf \dot{r}}_j(t))^2 
e^{i{\bf k}\cdot{\bf r}_j(t)}  
\!\!-\!\frac{1}{mV}\!\!\sum_{\bf k'}\!v_{k'} \!({\bf k}\!\cdot \!{\bf k'})   
\rho_{{\bf k}\!-\!{\bf k'}}(t)\rho_{\bf k'}(t) \!+\!\!\int_0^t 
\!\!\!dt'\gamma_{\bf k}(t\!-\!t'\!)\dot{\rho}_{\bf k}(t')  
\label{randomf1}  
\end{eqnarray}  
and,  
\begin{equation}   
\eta^{(I\!I)}_{\bf k}(t)=  
\ddot{\rho}_{\bf k}(t)+\Omega_k \rho_{\bf k}(t)+  
\int_0^t dt' \gamma_{\bf k}(t-t') \dot{\rho}_{\bf k}(t')  
\label{randomf2}  
\end{equation}  
  
\noindent  
For convenience of notation, we will refer from now on to
eq. (\ref{randomf1}) as representation I of the random force (or
explicit representation), and to eq. (\ref{randomf2}) as
representation II ( or implicit representation).  The requirement that
the two are equal, which is to say that the decomposition of the
residual force (\ref{expansion}) is valid, is clearly equivalent to
consider the equality between the two sides of equations
(\ref{der2rho2}), since the integral term, that we define for
simplicity,
\begin{equation}  
\Delta_{\bf k}(t)=\int_0^t \gamma_{\bf k}(t-t')   
\dot{\rho}_{\bf k}(t') dt'  
\label{integralterm}  
\end{equation}  
is equal in the two representations. Indeed, this arises from the fact  
that the memory function $\gamma_{\bf k}(t)$, being the correlation  
function of the noise following FDT (\ref{fdt}), must also be equal in  
the two representations.  
  
This will be only strictly valid if we could close exactly the
hierarchy illustrated above, which clearly can also be seen as
imposing constraints on the two sides of the density equations
(\ref{der2rho2}), or equivalently on the two representations of the
noise. Since this is a far from practical program to carry on, we can
try to identify those essential features of the system where
consistency should be enforced, in order to render the whole approach
simpler, but still satisfying the basic physical and mathematical
properties.

Consider now the two representations for the random noise  
(\ref{randomf1}) and (\ref{randomf2}). As noted above, the two should be  
equal to one another with respect to any averaging process we  
build.  A limited expression of this constraint is expressed if we  
ensure the validity of the equalities for any order $n$,  
\begin{equation}  
\langle \left(\prod_{i=1}^n \rho_{-{\bf k_i}}(t)\right)   
\eta^{(I)}_{\bf k}(t)\rangle =   
\langle \left(\prod_{i=1}^n \rho_{-{\bf k_i}}(t)\right)   
\eta^{(I\!I)}_{\bf k}(t)\rangle   
\label{ybg_n}  
\end{equation}  
This means that the two sides of equation (\ref{der2rho2}) will be
equal under projection onto density, density pairs, triplets and so
on. It must be also noted that since these are all equal time averages they
may be worked out exactly at any order and reduced to relations
between the density correlation functions, and the potential.  A clear
example of this was given by the YBG equation (\ref{ybg}),
corresponding simply to the case $n=1$ of the general condition
(\ref{ybg_n}).\\
A second example, that will have an important part in the following
discussion, is represented by the case where $n=2$, corresponding to
the ensuring of the equality of the two random force representations
under projection onto density pairs. Explicitly this can be written as,
\begin{equation}  
\langle \rho_{-{\bf p}}(t) \rho_{-{\bf q}}(t)   
\eta^{(I)}_{\bf k}(t) \rangle =  
\langle \rho_{-{\bf p}}(t) \rho_{-{\bf q}}(t)  
\eta^{(I\!I)}_{\bf k}(t)\rangle  
\label{ybg2implicit}  
\end{equation}  
Of course, the contribution on the two sides of the integral term  
$\Delta_{\bf k}(t)$ (\ref{integralterm}) cancels out, because we are  
assuming in principle the exactness of the equality of the two  
representations at any level.  Writing explicitly  
(\ref{ybg2implicit}), we have,  
\begin{eqnarray}  
\!\!\!\!\!\!\!\!\!\!\!\!&&-\frac{1}{mV}  
\sum_{\bf k'} ({\bf k}\cdot {\bf k'}) v_{k'}  
\langle \rho_{-{\bf p}}(t) \rho_{-{\bf q}}(t) \rho_{{\bf k}-{\bf k'}} (t)  
\rho_{\bf k'}(t)\rangle-
\frac{k^2}{\beta m} n c_k   
\langle \rho_{-{\bf p}}(t) \rho_{-{\bf q}}(t) \rho_{\bf k}(t)\rangle
\nonumber\\ \!\!\!\!\!\!\!\!\!\!\!\!&&=  
\langle \rho_{-{\bf p}}(t) \rho_{-{\bf q}}(t)   
\ddot{\rho}_{\bf k}(t)\rangle+\Omega_{\bf k}  
\langle \rho_{-{\bf p}}(t) \rho_{-{\bf q}}(t) \rho_{\bf k}(t)\rangle  
\label{constraint}  
\end{eqnarray}  
\noindent  
The terms $\langle \rho_{-{\bf p}}(t) \rho_{-{\bf q}}(t)   
\ddot{\rho}_{\bf k}(t)\rangle$ are easily calculated. Indeed,  
\begin{eqnarray} \!\!\!\! 
\langle \rho_{-{\bf p}}(t) \rho_{-{\bf q}}(t)   
\ddot{\rho}_{\bf k}(t)\rangle= -[\langle   
\dot{\rho}_{-{\bf p}}(t) \rho_{-{\bf q}}(t) \dot{\rho}_{\bf k}(t)\rangle  
+\langle \rho_{-{\bf p}}(t) \dot{\rho}_{-{\bf q}}(t)   
\dot{\rho}_{\bf k}(t)\rangle ] 
\label{coefficients1}  
\end{eqnarray}  
and,  
\begin{eqnarray}  
\langle   
\dot{\rho}_{-{\bf p}}(t) \rho_{-{\bf q}}(t) \dot{\rho}_{\bf k}(t)\rangle&=&  
\frac{{\bf p}\cdot {\bf k}}{\beta m}  
\delta_{{\bf p}+{\bf q},{\bf k}} N S_q\nonumber\\  
\langle \rho_{-{\bf p}}(t) \dot{\rho}_{-{\bf q}}(t)   
\dot{\rho}_{\bf k}(t)\rangle&=&\frac{{\bf q}\cdot {\bf k}}{\beta m}  
\delta_{{\bf p}+{\bf q},{\bf k}} N S_p  
\label{coefficients2}  
\end{eqnarray}  
Thus, rearranging equation (\ref{constraint}), considering also that $(1-n  
c_k)S_k=1$, we finally have,  
\begin{eqnarray}  
&&\!\!\!\!\!\!\!\!\!\!\!\!-\frac{1}{V}  
\sum_{\bf k'} ({\bf k}\cdot {\bf k'}) v_{k'}  
\langle \rho_{-{\bf p}}(t) \rho_{-({\bf k}-{\bf p})}(t)   
\rho_{{\bf k}-{\bf k'}} (t)  
\rho_{\bf k'}(t)\rangle- \frac{k^2}{\beta}  
\langle \rho_{-{\bf p}}(t) \rho_{-({\bf k}-{\bf p})}(t)   
\rho_{\bf k}(t)\rangle\nonumber\\  
&&\!\!\!\!\!\!\!\!\!\!\!\!
=-\frac{N}{\beta}\{k^2-{\bf k}\cdot {\bf p}\ n c_p    
-{\bf k}\cdot ({\bf k}-{\bf p})\ n c_{|{\bf k}-{\bf p}|}\} S_p   
S_{|{\bf k}-{\bf p}|}  
\label{ybg2}  
\end{eqnarray}  
\noindent  
This represents indeed a further relation involving the bare potential
and the static structure factor of the system, of the same type of the
YBG equation (\ref{ybg}), but of higher order, now also involving the
four density correlatione.  We note that, if one consider the case
${\bf p}=0$ in equation (\ref{ybg2}), this simply reduces to
YBG equation.
  
To sum up, equations (\ref{ybg_n}) for any $n$ constitute a minimal  
set of exact equations for equilibrium correlation functions that must  
be satisfied if Newtons equations are to be satisfied. The first of  
such equations is known as the YBG equation, and has been widely  
studied. We have earlier shown that this consistency relationship has  
the effect of modelling the collective part of the particle motion in  
an optimal manner. The second consistency relationship has not yet  
been the subject of interest in liquid state theory. It may be  
expected to play a role in ensuring that leading effects of phase  
relations between density variables are maintained, as well as some  
aspects of the dissipative effects of interactions that lead to  
decorrelation of the density variables.  
   
The set of equations (\ref{ybg_n}) will ensure the fidelity of the
Langevin process to the appropriate Newtonian dynamics on a single
time slice in the history of the system. Indeed, they do not contain
information about correlations between different time slices.

\section{Exact Calculation of the Memory Kernel}  
\label{sec:exact}  
\noindent  
Now we apply the FDT condition (\ref{fdt}) to calculate the  
memory kernel.  Following the definitions above, we have now two ways  
to proceed, by using the two equivalent representations for the random  
force $\eta_{\bf k}(t)$ (\ref{randomf1}),(\ref{randomf2}).  
  
Using the explicit form (\ref{randomf1}), which contains in it the  
bare potential $v_k$, it is straightforward to write an exact  
expression for the memory kernel applying FDT,  
\begin{eqnarray}   
\!\!\!\!\!\!\!\!\!\!\!\!\!\!\!\!\!\!\!\!\!\!\!\!\!\!\!\!\!\!\!\!&& \gamma^{(I)}_{\bf k}(t)=\frac{\beta m}{N k^2} \left[  
\left(\frac{k^2}{\beta m}\right)^2(n^2c_k^2-1)NS_k(t)+  
\langle \sum_{l,m}({\bf k}\cdot {\bf \dot{r}}_l(0))^2 e^{-i{\bf k}\cdot    
{\bf r}_l(0)} ({\bf k}\cdot {\bf \dot{r}}_m(t))^2 e^{i{\bf k}\cdot    
{\bf r}_m(t)} \rangle \right.   \nonumber\\    
\!\!\!\!\!\!\!\!\!\!\!\!\!\!\!\!\!\!\!\!\!\!\!\!\!\!\!\!\!\!\!\!&&+ \left.\frac{1}{(m V)^2}   
\sum_{\bf k'}\sum_{\bf k''}v_{k'}v_{k''}({\bf k}\cdot {\bf k'})   
(-{\bf k}\cdot {\bf k''}) 
\langle \rho_{-{\bf k}-{\bf k''}}(0) \rho_{\bf k''}(0)    
\rho_{{\bf k}-{\bf k'}}(t) \rho_{\bf k'}(t)\rangle \right.   \nonumber \\    
\!\!\!\!\!\!\!\!\!\!\!\!\!\!\!\!\!\!\!\!\!\!\!\!\!\!\!\!\!\!\!\!&&+ \left.\frac{n k^2}{\beta m^2 V} c_k \left\{   
\sum_{\bf k'} v_{k'}({\bf k}\cdot {\bf k'})\langle \rho_{-{\bf k}}(0)   
\rho_{{\bf k}-{\bf k'}}(t) \rho_{\bf k'}(t)\rangle 
+
\sum_{\bf k''} v_{k''}(-{\bf k}\cdot {\bf k''})\langle   
\rho_{-{\bf k}-{\bf k''}}(0) \rho_{\bf k''}(0)\rho_{\bf k}(t) 
\rangle \right\}   
\right.   \nonumber\\   
\!\!\!\!\!\!\!\!\!\!\!\!\!\!\!\!\!\!\!\!\!\!\!\!\!\!\!\!\!\!\!\!
&&- \left. \int_{0}^t dt' \gamma_{\bf k}(t-t')
\left\{\frac{k^2}{\beta m}n N c_k \dot{S}_k(t') 
+\frac{1}{mV}\sum_{\bf k'} v_{k'}   
(-{\bf k}\cdot {\bf k'})
\langle \rho_{-{\bf k}-{\bf k'}}(0) \rho_{\bf k'}(0)\dot{\rho}_{\bf k}(t')   
\rangle \right\}\right].   
\label{eq:gamma}   
\end{eqnarray}   
Thus, to explicitly evaluate the memory kernel, it is necessary to  
introduce some kind of approximations to evaluate the   
multiple density correlations appearing in (\ref{eq:gamma}).  \\
On the contrary, when using the implicit representation (II) for the  
random force (\ref{randomf2}), the formally exact expression for the  
memory kernel, would be,  
\begin{eqnarray}   
\!\!\!\!\!\!\!\!\!\!\!\!\!\!\!\!\!\!\!
\gamma^{(I\!I)}_{\bf k}(t)&=&\frac{\beta m}{N k^2} 
\left\{  
\langle \ddot{\rho}_{-{\bf k}}(0)\ddot{\rho}_{\bf k}(t)\rangle+  
\Omega_k [\langle \ddot{\rho}_{-{\bf k}}(0)\rho_{\bf k}(t) \rangle +  
\langle \rho_{-{\bf k}}(0)\ddot{\rho}_{\bf k}(t)\rangle]
\right. \nonumber\\ 
\!\!\!\!\!\!\!\!\!\!\!\!\!\!\!\!\!\!\!&+& \left.  
(\Omega_k)^2 \langle \rho_{-{\bf k}}(0)\rho_{\bf k}(t) \rangle   
+\int_{0}^t dt' \gamma_{\bf k}(t-t') \langle   
(\ddot{\rho}_{-{\bf k}}(0)+  
\Omega_k \rho_{-{\bf k}}(0)) \dot{\rho}_{\bf k}(t')\rangle \right\}  
\label{eq:gamma2}   
\end{eqnarray}   
\noindent  
This contains, for example, correlations at unequal times of the  
second derivatives of densities which are not known, even in an  
approximate form.  Thus, to pursue this second path, we must find a  
more suitable form for the random force to calculate its time  
correlation function.  This is precisely the route that has been used  
in the original derivation of the mode coupling theory (MCT)  
\cite{gotze}. Clearly this second possibility, despite the presence of these
intractable objects, does not contain explicitly the bare
potential. 
Of course, in an exact expression of the theory, we should have,  
\begin{equation}  
\gamma^{(I)}_{\bf k}(t)=\gamma^{(I\!I)}_{\bf k}(t)  
\end{equation}  
but we caution that different types of approximations made on the two  
expressions for the memory kernel may lead to inconsistent results.

\section{Representation I:   
Random Phase Approximation and Mode Coupling Theory}  
\label{sec:rep1}  
We now proceed to perform the simplest approximation to multiple
density correlations, to solve the YBG equation (\ref{ybg}) and at the
same time calculate the memory function (\ref{eq:gamma}) in
representation I. To do so, we consider the simplest case of a
Gaussian random process $\eta_{\bf k}(t)$. This implies, for example,
that the fourth moment of the force can be decomposed in terms of
second moments, as for example,
\begin{eqnarray}   
&&\langle \eta_{-{\bf k}}(t_1)\eta_{-{\bf k}}(t_2)\eta_{\bf k}(t_3)
\eta_{\bf k}(t_4)   
\rangle=\langle \eta_{-{\bf k}}(t_1)\eta_{-{\bf k}}(t_2)\rangle  
\langle \eta_{\bf k}(t_3) \eta_{\bf k}(t_4)\rangle \nonumber \\   
&&+\langle \eta_{-{\bf k}}(t_1)\eta_{\bf k}(t_3)\rangle  
\langle \eta_{-{\bf k}}(t_2)\eta_{\bf k}(t_4)\rangle+  
\langle \eta_{-{\bf k}}(t_1)\eta_{\bf k}(t_4)\rangle  
\langle \eta_{-{\bf k}}(t_2) \eta_{\bf k}(t_3) \rangle   
\label{momentum4}  
\end{eqnarray}  
Since we know that density variables are linear functions of the noise
from (\ref{laplacequations}), also densities are then Gaussian
variables. This means, for example, that the triplet correlation
function appearing in the YBG equation (\ref{ybg}) can be decomposed
as,
\begin{eqnarray}   
\langle \rho_{-{\bf k}}(t)\rho_{{\bf k}-{\bf k'}}(t)\rho_{\bf k'}(t) 
\rangle=   
\langle \rho_{-{\bf k}}(t)\rangle\langle \rho_{{\bf k}-{\bf k'}}(t) \rho_{\bf  
k'}(t)\rangle+ \nonumber \\   
\langle\rho_{{\bf k}-{\bf k'}}(t)\rangle\langle   
\rho_{-{\bf k}}(t)\rho_{\bf k'}(t)\rangle   
+\langle\rho_{\bf k'}(t)\rangle\langle\rho_{-{\bf k}}(t)  
\rho_{{\bf k}-{\bf k'}}(t) \rangle   
\label{tripletybg}  
\end{eqnarray}  
and by using the general result,  
\begin{equation}   
\langle \rho_{-{\bf p}}(t)\rho_{\bf q}(t) \rangle=N \delta_{{\bf  
p},{\bf q}} [S(q)+N \delta_{{\bf q},0}]   
\label{densitycorr2}  
\end{equation}  
we have,  
\begin{eqnarray}
\!\!\!\!\!\!\!\!\!\!\!\!\!\!\!\!\!\!\!\!\!\!\!\!\!\!\!\!\!\!\!\!\!\!\!\!\!\!\!
\!\!\!
\langle \rho_{-\!{\bf k}}(t)\rho_{{\bf k}\!-\!{\bf k'}}(t)
\rho_{\bf k'}(t)\rangle  
\!=\!N\delta_{{\bf k},0}[S(k')\!+\! N\delta_{{\bf k'},0}]
\!+\!N \delta_{{\bf k},{\bf k'}}[S(k)\!+\!N\delta_{{\bf k},0}]\!+\!N  
\delta_{{\bf k'},0}[S(k)\!+\! N\delta_{{\bf k},0}]   
\label{tripletybg2}  
\end{eqnarray}  
  
The result of inserting (\ref{tripletybg2}) into (\ref{ybg}) is that  
only terms where ${\bf k}={\bf k'}$ give a finite contribution in the  
sum on the right hand side, but this term has been already isolated  
from the rest of the sum in (\ref{ybg}).  Thus, this corresponds to a  
Random Phase Approximation (RPA) of the YBG equation. Therefore, we  
have that the structure factor is determined directly from the  
potential, via the relation,  
\begin{equation}    
c_{\bf k}=-\beta v_{\bf k}   
\label{msaapprox}  
\end{equation}  
\noindent  
Now, consistent with this approximation, we can also calculate the  
density averages for unequal times in (\ref{eq:gamma}), as for example,  
\begin{eqnarray}   
\!\!\!\!\!\!\!&&\langle \rho_{-{\bf k}-{\bf k'}}(0) \rho_{\bf k'}(0)   
\rho_{{\bf k}-{\bf k''}}(t) \rho_{\bf k''}(t)\rangle =   
\langle \rho_{-{\bf k}-{\bf k'}}(0) \rho_{\bf k'}(0)\rangle   
\langle \rho_{{\bf k}-{\bf k''}}(t) \rho_{\bf k''}(t)\rangle \nonumber \\  
\!\!\!\!\!\!\!&&+\langle   
\rho_{-{\bf k}-{\bf k'}}(0) \rho_{{\bf k}-{\bf k''}}(t) \rangle   
\langle\rho_{\bf k'}(0)\rho_{\bf k''}(t)\rangle+ \langle   
\rho_{-{\bf k}-{\bf k'}}(0)\rho_{\bf k''}(t)\rangle\langle   
\rho_{\bf k'}(0)\rho_{{\bf k}-{\bf k''}}(t) \rangle   
\label{quadrupla}  
\end{eqnarray}  
\noindent   
Also, to complete the calculation of the memory kernel, there is a  
second minor approximation that we need to introduce to evaluate the  
second term in the right hand side of (\ref{eq:gamma}). Thus, we  
perform the average over the velocities, even if now we are not  
dealing with equal time averages, and so neglecting the correlations  
in the single particle kinetic energy,  
\begin{eqnarray}   
\langle \sum_l\sum_m ({\bf k}\cdot {\bf \dot{r}}_l(0))^2 e^{-i{\bf k}
\cdot    
{\bf r}_l(0)} ({\bf k}\cdot {\bf \dot{r}}_m(t))^2 e^{i{\bf k}\cdot    
{\bf r}_m(t)} \rangle \approx \frac{k^4}{\beta^2 m^2} N S_k(t)    
\label{eq:approxv}   
\end{eqnarray}   
Thus, the Gaussian nature of the random process, and approximation
(\ref{eq:approxv}), allow to rewrite (\ref{eq:gamma}) as,
\begin{eqnarray}   
\!\!\!\!\!\!\!\!\!\!\!\gamma_{\bf k}(t)&=&\frac{n \beta}{m V k^2}   
\sum_{{\bf k'}\neq {\bf k}}\{v_{k'}^2({\bf k}\cdot {\bf k'})^2   
+ v_{k'}v_{k-k'}({\bf k}\cdot {\bf k'})({\bf k}\cdot ({\bf k} -{\bf k'}))   
\} S_{|{\bf k}-{\bf k'}|}(t) S_{k'}(t)  \nonumber \\   
\!\!\!\!\!\!\!\!\!\!\!&+& \frac{k^2 n^2}{\beta m }(c_k+\beta v_k)^2 S_k(t)- 
n (c_k+\beta v_k)\int_{0}^t dt'    
\gamma_{\bf k}(t-t')\partial_{t'}S_k(t')    
\label{eq:resultgamma}   
\end{eqnarray}   
At this point, we have to impose the YBG condition, consistent with  
Gaussian averages, that is the RPA condition (\ref{msaapprox}), and  
this leads precisely to the well-known expression given by the Mode  
Coupling Theory (MCT) for the memory kernel,  
\begin{eqnarray}  \!\!\!\!\!\!\!\!\! \!\!\!\!\!\!\!\!\! \!\!\!\!\!\!\!\!\!
\gamma_{\bf k}(t)^{^{MCT}} \!\!\!\!\!\!\!\!\!
= \frac{n}{\beta m  k^2 V}   
\sum_{{\bf k'}\neq {\bf k}}\{({\bf k}\cdot {\bf k'})^2   
c_{\bf k'}^2+    
({\bf k}\cdot {\bf k'})({\bf k}\cdot ({\bf k} -{\bf k'}))   
c_{\bf k'}   
c_{{\bf k} -{\bf k'}}\} S_{|{\bf k}-{\bf k'}|}(t) S_{k'}(t).   
\label{eq:memoryRPA}   
\end{eqnarray}   
\noindent  
It is interesting to note that in the Gaussian approximation, by
substituting the bare potential with the direct correlation function,
the contributions to the memory kernel given by the integral term
$\int_0^t \gamma_{\bf k}(t-t')\dot{\rho}_{\bf k}(t')$, present in the
noise are zero. If this were not the case, the expression for the
memory kernel would not be explicit any more, but it would be instead
a self-consistent relation. As it will be shown later, this will be
also valid if one calculates consistently in a Gaussian approximation
the memory kernel in representation II, projecting onto density pairs,
but leaving aside the integral term. On the contrary, in the original
derivation of MCT \cite{gotze}, using representation II, the
approximations made are not all consistent with each other, and in
that derivation, this term was a priori neglected, as explained in
detail in section \ref{sec:gotze}.
 
By introducing a factor $1/2$ in (\ref{eq:memoryRPA}), and calling  
${\bf k'}={\bf p}$ and ${\bf k}-{\bf k'}={\bf q}$, we can rewrite the  
memory kernel in the conventional MCT form, i.e.  
\begin{equation}   
\!\!\!\!\gamma_{\bf k}(t)^{^{MCT}}\!\!\!\!\!=\frac{n}{2\beta m V}
\!\!\!\!\!\!  
\sum_{\scriptsize \begin{array}{c} {\bf q}+{\bf p}={\bf k}\\   
{\bf q} \neq 0\end{array}}\!\!\!\!\!\!\!\!  
|({\bf q}c_{\bf q}+{\bf p}c_{\bf p}) \cdot e^L({\bf k})|^2 S_pS_q  
\Phi_p(t)\Phi_q(t)   
\label{memorymct}  
\end{equation}  
where $\Phi_k(t)$ is the normalised density correlator, for which the
equations of motion are now clearly,
\begin{equation}   
\ddot{\Phi}_{\bf k}(t)+\Omega_{\bf k}\Phi_{\bf k}(t)+  
\int_0^t \gamma^{^{MCT}}_{\bf k}(t-t') \dot{\Phi}_{\bf k}(t') dt' = 0  
\end{equation}  
  


\section{Projection onto Density Pairs}  
\label{sec:rep2}  
As we have discussed earlier, the formal exact expression of the
memory kernel in representation II (\ref{eq:gamma2}) does not
constitute a good starting point for making approximations and
developing a simple theory, because of the presence of the second time
derivatives of densities.  A way to overcome this problem is to
project the random force (\ref{randomf2}) onto the density pairs
subspace, as it was done in the original derivation of MCT
\cite{gotze}.  We, thus, leave unchanged the integral term
$\Delta_{\bf k}(t)$ (\ref{integralterm}), which appears in the
expression of the random force (\ref{randomf}), and in particular in
(\ref{randomf2}), and project the remaining part, which is clearly
equal to the minimum residual force ${\cal F}_{\bf k}(t)$ onto the
density pairs subspace. This choice follows naturally from the the YBG
condition, which states that the residual force is orthogonal to the
density at every time. 

For convenience, we define a {\it reduced} random force
$\tilde{\eta}_{\bf k}(t)$ that is simply the random force from which
we have subtracted the term $\Delta_{\bf k}(t)$ (\ref{integralterm}),
such as,
\begin{equation}  
\tilde{\eta}_{\bf k}(t)=\eta_{\bf k}(t)-\Delta_{\bf k}(t)  
\label{randomtilde}  
\end{equation}  
To be consistent, this reduced random force coincides with the
residual force ${\cal F}_{\bf k}(t)$, but we prefer to use this
notation at this stage to underline the fact that we are talking of a
stochastic process, even if mapped onto a deterministic one.  As for
the pure random force, also $\tilde{\eta}_{\bf k}(t)$ has of course
two representations. Now, following \cite{gotze}, we can rewrite the
random force in representation II as,
\begin{equation}  
\eta^{(I\!I)}_{\bf k}(t) = 
\sum_{l,m,p,s}\langle \rho_{-{\bf l}}(t)\rho_{-{\bf m}}(t)  
\tilde{\eta}_{\bf k}(t)^{(I\!I)}\rangle g(p,s,l,m) 
\rho_{\bf p}(t)\rho_{\bf s}(t)+ 
\Delta_{\bf k}(t)
\label{pairdecomposition}  
\end{equation}     
The coefficients of the expansion $\langle \rho_{-{\bf
l}}(t)\rho_{-{\bf m}}(t) \tilde{\eta}^{(I\!I)}_{\bf k}(t)\rangle$ can
be easily calculated for the first two terms in the expression of the
force (\ref{randomf2}). As we will show later, in the conventional MCT
derivation, the integral term is simply neglected in the calculation
of the memory kernel.  The unknown matrix $g(p,s,l,m)$ is the
normalisation matrix of the projection. In these matrices the bare
potential is now buried.  These are defined implicitly via the
relation,
\begin{equation}  
\!\sum_{l,m}\langle \rho_{-{\bf p}}(t)\rho_{-{\bf s}}(t)  
\rho_{\bf l}(t)\rho_{\bf m}(t) \rangle g(l,m,s',p') \!=\!  
\delta_{{\bf s},{\bf s'}} \delta_{{\bf p},{\bf p'}}  
\label{matrix}  
\end{equation}  
and it is clear from this that to explicitly evaluate $g(p,s,l,m)$
there is the need to evaluate four densities correlation functions,
and this can be done only by introducing some approximation for the
stochastic process. For the sake of simplicity, in the following we
neglect $\Delta_{\bf k}(t)$ but later we will show how it can easily
be included in the derivation, and how it might be important to do so.

The first important observation to make is that we can also think of
performing the same projection onto density pairs as in
(\ref{pairdecomposition}) for the representation I of the random force
(\ref{randomf1}). When this is done, it is easy to show that, at this
level, this can be reproduced exactly, by making use of the definition
of the normalisation matrix (\ref{matrix}), without the need to
calculate it explicitly. Let us show this in detail.
  
We want to calculate the coefficients of the projection for  
representation I of the reduced random force. We have,  
\begin{eqnarray} 
\langle \rho_{-{\bf l}}(t)\rho_{-{\bf m}}(t)   
\tilde{\eta}^{(I)}_{\bf k}(t)\rangle=-\frac{k^2}{\beta m} n c_k\langle   
\rho_{-{\bf l}}(t)\rho_{-{\bf m}}(t) \rho_{\bf k}(t)\rangle 
\nonumber\\ -\frac{1}{mV}  
\sum_{\bf k'} ({\bf k}\cdot {\bf k'}) v_{k'}  
\langle \rho_{-{\bf l}}(t)\rho_{-{\bf m}}(t)   
\rho_{{\bf k}-{\bf k'}}(t) \rho_{\bf k'}(t)\rangle  
\end{eqnarray}  
We note that these are static quantities, and thus the kinetic term in
(\ref{randomf1}) can be exactly rewritten in terms of density when
performing the averages, by separating out the velocity contribution.
Thus, we insert these coefficients in the projection definition  
(\ref{pairdecomposition}), and we have, omitting the time  
dependence since everything is at the same time $t$,  
\begin{eqnarray} \!\! \!\! \!\! \!\! \!\! \!\! \!\! \!\!\!\!\!\! \!\! \!\! 
\!\!   \!\! \!\! \!\! \!\! \!\! 
\tilde{\eta}^{(I)}_{\bf k}&=&  \! \! \! \! 
\sum_{l,m,p,s}\!\langle  \rho_{-{\bf l}}\rho_{-{\bf m}}  
\tilde{\eta}^{(I)}_{\bf k}\rangle g(p,s,l,m)  
\rho_{\bf p}\rho_{\bf s}
=-\frac{k^2}{\beta m} n c_k \sum_{p,s}\left\{\sum_{l,m}  
\langle \rho_{-{\bf l}}\rho_{-{\bf m}} \rho_{\bf k}\rangle   
g(p,s,l,m)\right\}   
\rho_{\bf p}\rho_{\bf s}  
\nonumber\\  \!\! \!\! \!\! \!\! \!\! \!\! \!\! \!\! \!\! \!\!  
\!\! \!\! \!\! \!\! \!\! &-&\frac{1}{mV}  
\sum_{\bf k'} ({\bf k}\cdot {\bf k'}) v_{k'}  
\sum_{p,s}\left\{\sum_{l,m}  
\langle \rho_{-{\bf l}}\rho_{-{\bf m}}   
\rho_{{\bf k}-{\bf k'}} \rho_{\bf k'}\rangle  
g(p,s,l,m)\right\} \rho_{\bf p}\rho_{\bf s}  
\label{projrep1}  
\end{eqnarray}  
We can thus evaluate explicitly the sums enclosed by graphs in  
(\ref{projrep1}) using (\ref{matrix}) as, 
\begin{eqnarray}  
&&\left\{\sum_{l,m}  
\langle \rho_{-{\bf l}}\rho_{-{\bf m}} \rho_{\bf k}\rangle   
g(p,s,l,m)\right\} = \frac{1}{N} \delta_{{\bf p},{\bf 0}}   
\delta_{{\bf s},{\bf k}}   
\label{elimina1}\\  
&&\left\{\sum_{l,m}  
\langle \rho_{-{\bf l}}\rho_{-{\bf m}}   
\rho_{{\bf k}-{\bf k'}} \rho_{\bf k'}\rangle  
g(p,s,l,m)\right\} = \delta_{{\bf p},{\bf k'}}   
\delta_{{\bf s},{\bf k}-{\bf k'}}   
\label{elimina2}  
\end{eqnarray}  
and by inserting these results into (\ref{projrep1}), remembering that  
$\rho_{k=0}\equiv N$, we find,  
\begin{eqnarray}  
\tilde{\eta}^{(I)}_{\bf k}(t)=   
-\frac{k^2}{\beta m} n c_k \rho_{\bf k}(t)-\frac{1}{mV}  
\sum_{\bf k'} ({\bf k}\cdot {\bf k'}) v_{k'}\rho_{{\bf k}-{\bf k'}} (t)  
\rho_{\bf k'}(t)  
\end{eqnarray}  
which is exactly the original expression we had for  
$\tilde{\eta}^{(I)}_{\bf k}(t)$ before projecting it.

This result tells us that, since the two representations of the noise
are intrinsically equal via the mapping between Newton and Langevin
equations, also the projection of the reduced force in representation
II onto density pairs in principle will be exact. Nonetheless, the
problem with it lies in the fact that it cannot be calculated
explicitly, because the definition (\ref{matrix}) cannot be applied
there to eliminate the unknown matrices $g$. Thus, differently from
the original work on MCT \cite{gotze}, where the projection onto pairs
was just chosen as the simplest approximation and not investigated
further, we can now say, through our knowledge of the Newtonian
representation which lies beneath oyr stochastic process, that it is
an exact step, if taken carefully, and it does not need further terms,
like for example one could think to project onto density triplets and
so on, as a better form of approximation.
  
Nonetheless, this projection must be treated carefully. Indeed, it  
tells us that we have chosen the right subspace to project onto, but  
it also impose a further constraint on our system.  As discussed  
previously, the equivalence between the two representations should,  
in principle, satisfy the constraints (\ref{ybg_n}) at any order  
$n$. Here, since the use of representation II involves the projection  
onto density pairs, it is explicitly required that our system  
satisfies at least the second order type of constraint, expressed by  
(\ref{ybg2}).  
  
Thus, this equation must be taken into account when using  
representation II of the random force to calculate the memory kernel,  
because the projection onto density pairs is now only exact when this  
constraint is satisfied.  
 
\subsection{Calculation of the Memory Kernel for the 
Projected Representation II}
We now proceed to calculate the memory function using the projection
onto density pairs of representation II.  Concerning the integral term
$\Delta_{\bf k}(t)$, we already know that its contribution is null
when we use the Gaussian approximation to evaluate the exact formula
for the memory kernel in representation I (\ref{eq:gamma}). In section
\ref{sec:gotze}, we will also reproduce the calculation of the kernel
as in the original MCT derivation \cite{gotze} and show that there the
term $\Delta_{\bf k}(t)$ is simply neglected.  Here, we will not
neglect it and show that it is possible to incorporate it in the
calculation, without rendering the calculation much harder.  Thus, as
illustrated in (\ref{pairdecomposition}), we do not project it onto
density pairs.

Using the explicit expression for the coefficients of the projection
for the second derivative term, given in (\ref{coefficients1}) and
(\ref{coefficients2}), we can write the `exact' expression for the
memory function, `exact' because it is only to be considered exact
when used in conjunction with the equation of constraint (\ref{ybg2}),
as discussed in the previous paragraph. Thus, applying (\ref{fdt}), 
and performing the sums that can be done explicitly so to isolate the
various types of density correlators, we have,
\begin{eqnarray}  
\!\!\!\!\!\!\!\!\!\!\!\!\!\!\!\!\!\!\!\!\!\!\!\!\!\!\!\!\!\!\!\!\!\!\!\!
&&\gamma_{\bf k}^{(I\!I)}(t)=\frac{N}{\beta m k^2}
{\large \sum}_{m,p,s}
{\large \sum}_{\scriptsize m'\!, p'\!,s'} [k^2\!-\!{\bf k}  
\!\cdot\!{\bf m'}nc_{m'}\!-\!{\bf k}\!\cdot\!({\bf k}\!-\!{\bf m'})nc_{k-m'}] 
[k^2\!-\!{\bf k} \!\cdot\!{\bf m}nc_{m}\!-\nonumber\\ 
\!\!\!\!\!\!\!\!\!\!\!\!\!\!\!\!\!\!\!\!\!\!\!\!\!\!\!\!\!\!\!\!\!\!\!\!&&\ \ 
\!{\bf k}\!\cdot\!({\bf k}\!-\!{\bf m})nc_{k-m}] 
S_{m'}S_{k-m'}S_{m}S_{k-m}g(p,s,k\!-\!m,m)g(k\!-\!m',m',p',s')  
\langle\rho_{-{\bf p'}}(0)\rho_{-{\bf s'}}(0)  
\rho_{\bf p}(t)\rho_{\bf s}(t) \rangle
\nonumber\\ 
\!\!\!\!\!\!\!\!\!\!\!\!\!\!\!\!\!\!\!\!\!\!\!\!\!\!\!\!\!\!\!\!\!\!\!\!
&&-\frac{1}{\beta m S_k} {\large \sum}_{\scriptsize m,p,s} 
[k^2\!-\!{\bf k} \!\cdot\!{\bf m}nc_{m}\!-
\!{\bf k}\!\cdot\!({\bf k}\!-\!{\bf m})nc_{k-m}] 
S_{m}S_{k-m}g(p,s,k-m,m)\langle\rho_{-{\bf k}}(0) 
\rho_{\bf p}(t)\rho_{\bf s}(t) \rangle \nonumber\\
\!\!\!\!\!\!\!\!\!\!\!\!\!\!\!\!\!\!\!\!\!\!\!\!\!\!\!\!\!\!\!\!\!\!\!\!
&&-\frac{1}{\beta m S_k} 
{\large \sum}_{\scriptsize m'\!,p'\!,s'}[k^2\!-\!{\bf k}  
\!\cdot\!{\bf m'}nc_{m'}\!-\!{\bf k}\!\cdot\!({\bf k}\!-\!{\bf m'})nc_{k-m'}]
S_{m'}S_{k-m'}g(k\!-\!m',m',p',s')
\nonumber\\ 
\!\!\!\!\!\!\!\!\!\!\!\!\!\!\!\!\!\!\!\!\!\!\!\!\!\!\!\!\!\!\!\!\!\!\!\!&&
\langle\rho_{-{\bf p'}}(0)\rho_{-{\bf s'}}(0)  
\rho_{\bf k}(t) \rangle  - 
\frac{1}{k^2} {\large \sum}_{\scriptsize m'\!,p'\!,s'}[k^2\!-\!{\bf k}  
\!\cdot\!{\bf m'}nc_{m'}\!-\!{\bf k}\!\cdot\!({\bf k}\!-\!{\bf m'})nc_{k-m'}]  
S_{m'}S_{k-m'} g(k\!-\!m',m',p',s')\nonumber\\
\!\!\!\!\!\!\!\!\!\!\!\!\!\!\!\!\!\!\!\!\!\!\!\!\!\!\!\!\!\!\!\!\!\!\!\!&&
\int_0^t\!\!\!dt'  
\gamma_{\bf k}(t\!-\!t')\langle \rho_{-{\bf p'}}(0)\rho_{-{\bf s'}}(0)  
\dot{\rho}_{\bf k}(t')\rangle +
\frac{k^2}{\beta m S_k^2} S_k(t)+
\frac{1}{N S_k}\int_0^t \!\!\!dt'\gamma_{\bf k}(t-t')  
\langle \rho_{-{\bf k}}(0)  
\dot{\rho}_{\bf k}(t')\rangle 
\label{gammaIIexact2}  
\end{eqnarray}  
This form has now the same structure as expression (\ref{eq:gamma})
for representation I. In other words, it is explicitly visible that it
contains the time-dependent 4-point correlation $\langle\rho_{-{\bf
p'}}(0)\rho_{-{\bf s'}}(0) \rho_{\bf p}(t)\rho_{\bf s}(t) \rangle$, as
well as triplets and pairs time-dependent correlations. Nonetheless,
now the coupling coefficients do not contain the bare interactions,
but already we can see that they are composed of static quantities,
such as the direct correlation function, although clearly the
unknown matrix $g$ still needs to be treated.

\subsection{The Gaussian case}  
\label{subsec:rep2gaussian}  
  
In this section we want to apply the simple Gaussian approximation for
the random process that we used to calculate the memory kernel in
representation I, to the case of projected representation II. The
first equation to study is of course (\ref{ybg2}), because this is the
one that guarantees the consistency of the whole scheme. Thus,
recalling that for a gaussian noise multiple averages can be
decomposed as in (\ref{momentum4}), the constraint (\ref{ybg2})
becomes in this simple case,
\begin{eqnarray}  
&&-\frac{1}{V}\{  
{\bf k}\cdot {\bf p}v_p +  
{\bf k}\cdot ({\bf k}-{\bf p})v_{|{\bf k}-{\bf p}|}\}N^2 S_p   
S_{|{\bf k}-{\bf p}|}-\frac{k^2}{\beta}\{\delta_{{\bf p},0}  
+\delta_{{\bf p},{\bf k}}\}N^2 S_k=\nonumber\\  
&&-\frac{N}{\beta}\{k^2-{\bf k}\cdot {\bf p}\ n c_p    
-{\bf k}\cdot ({\bf k}-{\bf p})\ n c_{|{\bf k}-{\bf p}|}\} S_p   
S_{|{\bf k}-{\bf p}|}  
\label{ybg2gaussian}  
\end{eqnarray}  
Now using the condition $c_k=-\beta v_k$, that we found applying the
gaussian approximation to the YBG equation, we have cancellation of
some terms, but then this would imply, $\{\delta_{{\bf p},0}
+\delta_{{\bf p},{\bf k}}\}N S_k=S_p S_{|{\bf k}-{\bf p}|} $ and this
is clearly not an equality valid for all ${\bf k}$ and ${\bf p}$, but
only for the special cases ${\bf k}={\bf p}$, or ${\bf p}=0$.  Thus,
we conclude that equation (\ref{ybg2}) cannot be satisfied by a simple
gaussian approximation. This, indeed, implies that using
representation II, projected onto density pairs, and proceeding with a
pure gaussian approximation to calculate the memory kernel, the result
found for it will not be the same as the one found with representation
I (\ref{eq:memoryRPA}), because the relation of constraint
(\ref{ybg2}) would be violated in this way.  As a consequence of this,
we have that, to use representation II, one must go a step further in
the approximation of the noise than with representation I, where the
simple gaussianity of the noise is sufficient to develop a consistent
theory. This comes from the fact that representation I is
intrinsically exact, while representation II is not, with respect to
the projection onto pairs, and thus, using the latter, one must be
more careful.\\
To evaluate explicitly (\ref{gammaIIexact2}), we need an expression
for the normalisation matrix $g$. Using again gaussian properties of
the noise to decompose density correlations, we have that
(\ref{matrix}) gives,
\begin{equation}  
g(s,p,p',s')\approx 
\frac{\delta_{{\bf p},{\bf p'}}\delta_{{\bf s},{\bf s'}}}  
{2 N^2 S_s S_p} 
\label{gaussiang} 
\end{equation}  
Using this result, and applying systematically gaussian decomposition  
of averages at equal and unequal times, we have the following result  
for the memory function,  
\begin{eqnarray}  \!\!\!\!\!\!\!\!\!\!\!\!\!\!\!\!\!\!
\gamma^{(I\!I)\ Gauss}_{\bf k}(t)&=& 
\!\!\!\frac{1}{2 \beta m k^2 N}  
\sum_p [k^2-{\bf k}\cdot {\bf p} \ n c_p -{\bf k}  
\cdot ({\bf k}-{\bf p})\ n c_{|{\bf k}-{\bf p}|}]^2 
S_p(t) S_{|{\bf k}-{\bf p}|}(t)  \nonumber \\ 
\!\!\!\!\!\!\!\!\!\!\!\!\!\!\!\!\!\!&-&\!\!\!
\frac{k^2}{\beta m S_k} (1-n c_k) S_k(t)-
\left[(1-nc_k)-\frac{1}{S_k}\right]  
\int_0^t\gamma_{\bf k}(t-t')\dot{S}_k(t')
\label{gammagaussII}  
\end{eqnarray}
The first thing to note here is that the contribution, given by the
integral term, vanishes because $[(1-nc_k)-1/S_k]=0$, and this
coincides with what we have found in representation I in gaussian
approximation.  Also, we can isolate from the sum the terms for which
$m$ is singular ($m=0$ and $m=k$), and, recalling equation
(\ref{memorymct}), we can rewrite (\ref{gammagaussII}) as,
\begin{eqnarray}\!\!\!\!\!\!\!\!\!\!\!\!\!\!\!\!\!\!
\gamma_{\bf k}(t)=
\gamma_{\bf k}(t)^{^{MCT}}\!\!\!\!\!\!+  
\!\frac{1}{2 \beta m k^2 N}\!\!
\sum_{{\bf p}\neq 0,{\bf k}} \!\![k^4\!\!-\!2{\bf k}\!\cdot\! {\bf p} n c_p   
\!\!-\!2{\bf k}\!\cdot\! ({\bf k}\!-\!{\bf p})n c_{|{\bf k}-{\bf p}|}]  
S_p(t) S_{|{\bf k}-{\bf p}|}(t)  
\end{eqnarray}  
Thus, clearly, this way of proceeding does not reproduce MCT result
and the reason of this lies in the fact that the two representations
for the noise (\ref{randomf1}), (\ref{randomf2}) are not equivalent
under projection onto density pairs using a simple Gaussian
approximation, because, as shown above (\ref{ybg2}) is not satisfied,
except for the trivial wave-vectors cases.  This means that further
investigation of the theory must involve a new type of approximation
for multiple density averages, that would satisfy firstly equation
(\ref{ybg2}).


\subsection{The STLS Approximation: a step beyond Gaussianity}  
\label{subsec:rep2singwi}  
\noindent  
We want to illustrate here a step in the direction of improving the
gaussian approximation for density averages.  Indeed, turning to
consider other known approximations in literature, it may be possible
to satisfy the constraint (\ref{ybg2}) for a larger number of
wave-vectors than the simple gaussian one.  The example we will refer
to is known as the Singwi-Tosi-Land-Sjolander approximation
(STLS) \cite{stls}.  What we show here does not lead to a definitive
result for corrections to MCT, but it is illustrative of the issues
involved.
  
We can consider now for example the triple average in the YBG equation  
(\ref{ybg}). Recalling what happened with the Gaussian approximation,  
we can now think that in this new scheme, we will have, as dominant  
contributions to the sum, not only the Gaussian terms but also some  
corrections to them, that would be most meaningful if they in some way  
include the gaussian terms, so that the new scheme is actually an  
improvement of the old one. To say this mathematically, we can write  
explicitly,  
\begin{eqnarray}   
&&\langle \rho_{-{\bf k}}(t)\rho_{{\bf k}-{\bf k'}}(t)
\rho_{\bf k'}(t)\rangle  
=N\{\delta_{{\bf k},0}+{\cal C}_1\}[S_{k'}+ N\delta_{{\bf k'},0}]
\nonumber\\  &&+N  
\{\delta_{{\bf k},{\bf k'}}+{\cal C}_2\}[S_k+N\delta_{{\bf k},0}]+N  
\{\delta_{{\bf k'},0}+{\cal C}_3\}[S_k+ N\delta_{{\bf k},0}]   
\label{triplet+corrections}  
\end{eqnarray}  
where ${\cal C}_{1,2,3}$ are the three corrections to the Gaussian  
terms, that are simply $\delta$-functions.  Thus, we are still assuming  
that the triplet correlation function decomposes into three separate  
contributions, as for the gaussian variables, but now these  
contributions will be more complicated.  
Following the literature \cite{stls}, the STLS consists in assuming
the following,
\begin{eqnarray}   \!\!\!\!\!\!\!\!\!\!
&&\rho_{{\bf k}-{\bf k'}}(t)\rho_{\bf k'}(t)=\sum_n e^{i({\bf k}-{\bf  
k'})\cdot {\bf r}_n(t)}\sum_{n'}e^{i{\bf k'}\cdot {\bf r}_{n'}(t)}  
=\sum_{n'}e^{i{\bf k}\cdot   
{\bf r}_{n'}(t)} \sum_n e^{i({\bf k}-{\bf k'})  
\cdot ({\bf r}_n(t)-{\bf r}_{n'}(t))}\nonumber\\  \!\!\!\!\!\!\!\!\!\!
&&\approx\sum_{n'}e^{i{\bf k}\cdot   
{\bf r}_{n'}(t)} \langle \sum_n e^{i({\bf k}-{\bf k'})\cdot   
({\bf r}_n(t)-{\bf r}_{n'}(t))}\rangle =\rho_{\bf k}(t)  
[S_{|{\bf k}-{\bf k'}|}+N\delta_{{\bf k},{\bf k'}}]   
\label{singwi}  
\end{eqnarray}  
where the approximation made consists of replacing the sum over $n$ on  
the second line of (\ref{singwi}) by its average, as it was  
independent on $n'$.  
  
If we apply this to the triplet correlation function above, we have,  
\begin{eqnarray}   
&&\langle \rho_{-{\bf k}}(t)
\rho_{{\bf k}-{\bf k'}}(t)\rho_{\bf k'}(t)\rangle =
N\{\delta_{{\bf k},0}+S_k\}[S_{k'}+ N\delta_{{\bf k'},0}]+\nonumber\\  
&&+N  \{\delta_{{\bf k},{\bf k'}}+S_{|{\bf k}-{\bf k'}|}\}  
[S_k+N\delta_{{\bf k},0}]+N  
\{\delta_{{\bf k'},0}+S_{k'}\}[S_k+ N\delta_{{\bf k},0}]   
\label{tripletsingwi}  
\end{eqnarray}  
and the corrections ${\cal C}_{1,2,3}$ are simply the static structure
factors, with wavenumber corresponding to the respective
$\delta$-function.  A physical picture of what this represents is
straightforward. Indeed, the gaussian decomposition of averages simply
means that the densities at three points are only coupled two by two
in all possible combinations, assuming that the third one is
infinitely far from the others and thus, this is averaged out. The new
STLS approach consists in accounting for some aspects of the
correlations of the third particle.  This is the reason why we now need
two structure factors to describe our triplet correlations, while in
the gaussian treatment we needed only one.
  
Now we apply the approximation (\ref{tripletsingwi}) to the YBG  
equation (\ref{ybg}), thus obtaining a new solution for the direct  
correlation function of the system, i.e.  
\begin{equation}   
c_k=-\beta v_k- \frac{\beta}{N k^2}\sum_{{\bf k'}\neq{\bf k}}v_{k'}   
({\bf k}\cdot{\bf k'}) S(|{\bf k}-{\bf k'}|)  
\label{ybgsingwi}   
\end{equation}  
where again only the contribution driven by the ${\bf k}={\bf k'}$  
$\delta$-function survives, because the other two can be both shown to  
have spherical symmetry and thus, their integral is null.  The result  
(\ref{ybgsingwi}) clearly still contains the information of the RPA in  
the first term, but adds to it a new term to be taken into account for  
the calculation of $S(k)$ from the potential. \\ 
The next step now is to see what happens when applying the STLS
approximation to the second equation of constraint (\ref{ybg2}). We
have seen that the Gaussian approximation would satisfy this condition
only for two singular wavevectors.  Using the STLS approximation for
the quadruplet and triplet correlations on the left hand side and
(\ref{ybgsingwi}) for the direct correlation function on the right
hand side, (\ref{ybg2}) becomes
\begin{eqnarray}\!\!\!\!\!\!\!\!\!\!\!\!\!\!\!\!\!\!\!\!\!
&&\sum_{\bf k'} v_{k'}({\bf k}\cdot {\bf k'}) [S_k S_p S_{k'}+  
(S_{|{\bf k'}-{\bf p}|}+S_{|{\bf k}-{\bf k'}-{\bf p}|})  
S_p S_{|{\bf k}-{\bf p}|}]=\nonumber\\  
\!\!\!\!\!\! \!\!\!\!\!\!\!\!\!\!\!\!\!\!\!
&&\sum_{\bf k'} v_{k'} \left[\frac{({\bf k}\cdot{\bf p})  
({\bf k'}\cdot {\bf p}) }{p^2}S_{|{\bf k'}-{\bf p}|}
+\frac{({\bf k}\cdot({\bf k}-{\bf p}))  
({\bf k'}\cdot ({\bf k}-{\bf p}))}{(|{\bf k}-{\bf p}|)^2}  
S_{|{\bf k}-{\bf k'}-{\bf p}|}\right]S_p S_{|{\bf k}-{\bf p}|}  
\end{eqnarray}  
The first of the three terms on the left hand side gives no  
contribution due to spherical symmetry. Thus, to satisfy the  
constraint (\ref{ybg2}) in the STLS approximation, we have to satisfy  
the general condition for certain wave-vectors ${\bf p}$,   
equivalently ${\bf k}-{\bf p}$,  
\begin{equation}  
({\bf k}\cdot {\bf k'})=\frac{({\bf k}\cdot{\bf p})  
({\bf k'}\cdot {\bf p})}{p^2}\nonumber\\  
\end{equation}  
This implies the geometrical condition
$\cos{\theta_{k,k'}}\!=\!\cos{\theta_{k,p}}\cos{\theta_{k',p}}$, which
is equivalent to
$\sin{\theta_{k,p}}\sin{\theta_{k',p}}\cos{\phi_{k,k'}}=0$.  Being
here ${\bf k'}$ arbitrary, the only general condition that can satisfy
this constraint is that ${\bf k}$ and ${\bf p}$ are parallel, plus of
course of both of them being zero.  This is certainly an enlarged
number of wave-vectors over which STLS satisfies the second order
constraint (\ref{ybg2}) with respect to the only two suitable values
for the gaussian approximation.  This type of scheme would possibly
provide a better way of proceeding in calculation of the memory
kernel, using the projection onto density pairs of representation
II. A problem, in doing so, would be how to extend correctly the STLS
approximation (\ref{singwi}) to unequal time correlations, and if then
would then be possible to close the theory, by eliminating the bare
potential, for example using the result of the YBG (\ref{ybgsingwi}).

These questions do not appear to have an easy answer, as well as the
challenge of satisfying the constraint (\ref{ybg2}) for any
wave-vector seems quite hard. That is why future work will be
addressed to propose an alternative, in principle self-consistently
determined, way to go beyond the gaussian closure, as well as to
renormalize the bare interactions.

  
\section{Comparison of our formalism with the original derivation of MCT} 
\label{sec:gotze} 
\noindent  
Here, we want to link our formalism the traditional derivation of MCT
\cite{gotze}, based on the projector operator formalism. In this way,
we will be able to show in more detail the various steps that were
originally taken, and compare them to our work.
  
Thus, we analyze the original derivation, starting form when equations
of motion are written for the density and current correlators,
following the Zwanzig-Mori (ZM) formalism, where the variables which
define the projection space are, indeed, the density $\rho_{\bf k}$
and the longitudinal current $\j_{\bf k}$ \cite{gotze}.  In the
subspace spanned by these variables, the projection operators are
defined as ${\cal P}$ being the projector onto the subspace itself and
${\cal Q}=1-{\cal P}$ the orthogonal projector. Thus, translating the
notation in \cite{gotze} into our notation, we have the following
equations of motion,
\begin{eqnarray}  \!\!\!\!\!\!\!\!\!\!\!\!\!\!\!\!\!\!\!\!
\left(\matrix{ \dot{\rho}_{\bf k}(t)\cr  
\dot{\j}_{\bf k}(t)\cr}  
\right)  
\!- \!i  
\left(\matrix{0&k\cr  
\Omega_k/k&0\cr}  
\right)  
\left(\matrix{{\rho}_{\bf k}(t)\cr  
{\j}_{\bf k}(t)\cr}  
\right)\!+\!
\int_0^t  \left(\matrix{0&0\cr  
0&\gamma_{\bf k}(t-t')\cr}  
\right)  
\left(\matrix{{\rho}_{\bf k}(t')\cr  
{\j}_{\bf k}(t')\cr}  
\right)\!= \! 
\left(\matrix{  
 0 \cr  
 R_{\bf k}(t)\cr}  
\right)  
\label{zmeqs}  
\end{eqnarray}  
that combined together give equation (\ref{langevin}), where $ R_{\bf
k}(t)$ is the random force which was indicated as $\eta_{\bf k}(t)$ in
our notation. There are a few things to specify regarding
(\ref{zmeqs}). Firstly, in the ZM formalism, the random force at the
initial time $t=0$, is defined for a general variable ${\cal A}$ as
$R_{\huge {\cal A}(0)}={\cal Q}\dot{\cal A}$, and so it is clear that
${\cal Q} \dot{\rho}_{\bf k}=0$ and the only non-vanishing random
force is related to the current. Thus we have,
\begin{equation}  
R_{\bf k}(0)={\cal Q} \dot{\j}_{\bf k}(0)=  
-i/k[\ddot{\rho}_{\bf k}(0)+\Omega_k   
\rho_{\bf k}(0)]  
\end{equation}  
Also, in this formalism, we know that the random force   
evolves in time not with the Liouvillian operator ${\cal L}$, but with a  
particular projection of it, and thus,  
\begin{equation}  
R_{\bf k}(t)=\exp{(i{\cal QLQ}t)}  
R_{\bf k}(0).  
\label{exactforce}  
\end{equation}  
Nonetheless, in the MCT approach, the evolution of the random force is  
approximated with the Liouvillian evolution in time, i.e.  with  
$e^{i{\cal L}t} R_{\bf k}(0)$.  Thus, the approximation corresponds to  
neglect the difference between $e^{i{\cal QLQ}t} R_{\bf k}(0)$ and  
$e^{i{\cal L}t} R_{\bf k}(0)$ i.e. 
\begin{equation} R_{\bf k}(t) \approx  
-\frac{i}{k}[\ddot{\rho}_{\bf k}(t)+\Omega_k \rho_{\bf k}(t)].  
\label{force}  
\end{equation}    
In our derivation, using representation II, the true random force
$\eta_{\bf k}(t)$ is given in (\ref{randomf2}), and it is clear that,
apart from the factor $-i/k$ that comes from using as a variable
$\j_{\bf k}(t)$ instead of $\dot{\rho}_{\bf k}(t)$, the two being
related via the relation $\dot{\rho}_{\bf k}(t)=ik\j_{\bf k}(t)$, this
is the same as the exact $R_{\bf k}(t)$ (\ref{exactforce}).  The
comparison between (\ref{randomf2}) and (\ref{force}) allows us to
give an exact expression for the difference between the true random
force and the MCT approximated one. Indeed, it turns out to be the
integral term $\Delta_{\bf k}(t)$, defined in (\ref{integralterm}).
We have shown how to include $\Delta_{\bf k}(t)$ in the equations for
the memory kernel. We have also shown that the contribution to the
memory kernel coming from $\Delta_{\bf k}(t)$ is strictly zero in RPA,
in both representations.
  
Thus, the starting point for the original derivation of the memory
kernel in MCT, is equation (\ref{gammaIIexact2}), in which one neglects
the integral term.  The approximations that are then made are the
following. Firstly, it is chosen to perform a kind of a gaussian
approximation for the four-point density average, which means that
still the quadruple averages factorises into averages of pairs, but
not all the possible combinations are considered, i.e.
\begin{eqnarray} \!\!\!\!\!\!\!\!\!\!\!\!\!\!\!\!\!\!\!\!  
\langle \rho_{-{\bf p'}}(0)\rho_{-{\bf s'}}(0)
\rho_{\bf p}(t)\rho_{\bf s}(t)   
\rangle \approx \langle \rho_{-{\bf p'}}(0)\rho_{\bf p}(t)\rangle  
\langle \rho_{-{\bf s'}}(0)\rho_{\bf s}(t)\rangle 
N^2 S_p(t) S_s(t) \delta_{{\bf p},{\bf p'}} \delta_{{\bf s},{\bf s'}}  
\label{gaussbla}  
\end{eqnarray}  
The same approach is followed to evaluate the normalization matrix
$g$, using definition (\ref{matrix}), i.e.  $g(l,m,p,s) \approx
\delta_{{\bf l},{\bf s}} \delta_{{\bf m},{\bf p}}/ N^2 S_p S_s$. These
types of simplified gaussian approximations can be substituted by the
conventional ones that we have used, like for example
(\ref{quadrupla}), without changing the final result for the MCT
memory kernel (\ref{memorymct}), because of the symmetry that this has
with respect to the wave-vectors ${\bf p}$ and ${\bf q}$.  Regarding,
instead, the triple averages appearing in the coefficients, they can
be exactly written as, for example,
\begin{equation}  
\langle \rho_{-{\bf l}}(t) \rho_{-{\bf m}}(t) \rho_{\bf k}(t)\rangle=  
N \delta_{{\bf l}+{\bf m},{\bf k}} S_l S_m S_k (1+n c_3)  
\label{c3}  
\end{equation}  
where $c_3$ indicates the direct correlation function for
triplets. Thus, based on the observation that for hard spheres systems
it has been shown that the contribution given by $c_3$ in (\ref{c3})
is very small, this is neglected. Also, if one wanted to include them
in the calculation, they can be easily calculated from
simulations. Thus, this approximations for the triplet density
correlations appears to be more advanced than the simple decomposition
of it, generated by a gaussian noise choice, as in
(\ref{tripletybg}). Nonetheless, this breaks the full consistency of
the scheme of approximations in developing the theory.
  
Eq. (\ref{c3}), but where the triplet direct correlation function is
neglected, leads to a crucial simplification of the coefficients of
the projection of the noise. Indeed, it implies a cancellation of the
triplet term with the term in $k^2$. Thus, one can see that this, in
addition to the gaussian type of approximation for the quadruple
correlations and the normalisation matrix discussed above, leads very
easily to the well known expression for $\gamma^{MCT}_{\bf k}(t)$
given in (\ref{memorymct}).
  
Now let us analyze what the inclusion of the integral term
$\Delta_{\bf k}(t)$ would produce in the present scheme. To calculate
the full memory kernel we are left to evaluate the last term in
(\ref{gammaIIexact2}). Using the approximations just discussed,
i.e. gaussian averages for 4-point correlations and (\ref{c3}) for the
triplets neglecting $c_3$, we have that the last term reduces to,
\begin{eqnarray}  
\!\!\!\!\!\!\!\!\!\!\!\!\!\!\!\!\!\!\!\!\!\!\!\!\!\!
&&\frac{\beta m}{N k^2}  \ {\large \sum}_{\scriptsize l',m',p',s'}  
\left\{-\frac{N}{\beta m}[k^2-{\bf k}  
\cdot{\bf m'}nc_{m'}- 
{\bf k}\cdot({\bf k}-{\bf m'})nc_{k-m'}]S_{m'}S_{q-m'}  
\delta_{l'+m',k} \right. \nonumber\\ 
\!\!\!\!\!\!\!\!\!\!\!\!\!\!\!\!\!\!\!\!\!\!\!\!\!\!&&\left.  
+\frac{k^2}{\beta m S_k} \langle \rho_{-{\bf k}}(0)\rho_{\bf l'}(0)  
\rho_{\bf m'}(0)\rangle \right\}g(l',m',p',s') 
\int_0^t \!dt' \gamma_{\bf k}(t-t')  
\langle \rho_{-{\bf p'}}(0)\rho_{-{\bf s'}}(0)  
\dot{\rho}_{\bf k}(t')\rangle \nonumber\\   
\!\!\!\!\!\!\!\!\!\!\!\!\!\!\!\!\!\!\!\!\!\!\!\!\!\!&&=
\!\frac{\beta m}{N^3 k^2}\!\sum_{m'}[{\bf k}  
\!\cdot\!{\bf m'}nc_{m'}\!+\!{\bf k}\!
\cdot\!({\bf k}\!-\!{\bf m'})nc_{k-m'}]\!
\int_0^t\!\!\!dt'\gamma_{\bf k}(t\!-\!t')\langle \rho_{-{\bf m'}}(0)  
\rho_{-{\bf k}+{\bf m'}}(0)  
\dot{\rho}_{\bf k}(t')\rangle   
\end{eqnarray}  
Thus, there is the need to calculate the dynamical average $\langle
\rho_{-{\bf m'}}(0)\rho_{-({\bf k}{\bf m'})}(0)\dot{\rho}_{\bf
  k}(t')\rangle$ to evaluate the contribution of this term. We could
use either the Gaussian approximation, as for the quadruple average
(\ref{gaussbla}), or a generalization for unequal times of the triplet
expression (\ref{c3}).  Nonetheless, in both cases, this contribution
does not appear to be zero, as in purely gaussian treatments, and
moreover, it requires the introduction of another approximation,
somewhat consistent with the others that have been made previously. In
any case, it would be certainly different from zero, and its simple
neglection, without a consistent justification of it, is clearly
unsatisfying.

  
\section*{Conclusions}  
  
In this paper we have attempted to provide an alternative way to
derive a mode-coupling theory.  We have shown how Newton's equations
can be represented by trial equations, and how these may be
approximated.  The approach starting from Newton's equations allows us
to derive explicit and exact expressions for the the memory function
and for the noise without the requirement of projection operators.It
thereby provides an alternative view that may lead to new insights.
The resulting equations, solved under the assumption that the noise,
and consequently the density fluctuations of the liquid, is gaussian
distributed, are equivalent to the random-phase-approximation for the
static structure factor and to the well known ideal mode coupling
theory (MCT) equations for the dynamics.  This suggests that MCT is a
theory of fluid dynamics that becomes exact in a mean-field limit.
This possibility was suggested some time ago on the basis of the
analogies between the equations describing the schematic MCT models
and the the dynamics of the order parameter in disordered $p$-spin
models, solved under strict mean-field approximation
\cite{thirumalai,crisanti,kurchan}.
 
We have also investigated in detail the basic meaning of the less
controlled steps intrinsic in the derivation of conventional
MCT \cite{gotze}.  One important point is the demonstration that, if
the projection of the noise over the density pairs could be carried
out without approximations, the resulting expression would be exact.
The evaluation of the quadruplet correlation functions under gaussian
statistics, an approximation employed in the evaluation of the memory
function when projected in the density pair subspace constitutes, in
conventional MCT, the step equivalent to the RPA approximation.
Hence, our approach suggests that the conventional MCT is exact in the
RPA limit. This offers the possibility of improving the theory by
improving the approximation of the four-point correlation functions in
the same way as one improves the static structure factors on moving
beyond RPA.  We have illustrated an example of this by using the STLS
approximation (\ref{singwi}), which leads to a different closure for
the YBG equation (\ref{ybgsingwi}), certainly more advanced than the
simple RPA. 

The implication of this is clearly not that ideal MCT is a theory
which can be applied only in the weak-coupling limit, where RPA
becomes exact.  In fact, the RPA limit that we have studied may in
principle be corrected in two generic manners. Firstly, the treatment
of the static correlations may be improved by use of a better direct
correlation function than that implied by the consistent theory.  On
the other hand, Kawasaki \cite{kawasaki} has presented an interesting
derivation of MCT based on the (quadratic) density functional
Ramakrishnan-Yussouf free energy of a liquid, where the effective
interaction between the density pairs is exactly $- c(k) /\beta$,
thereby accomplishing the limited form of renormalisation implied by
RPA.  In addressing this issue we note that the comparison between the
MCT prediction and hard-core type experimental
\cite{goetzepisa,vanmegen} or numerical \cite{koblj} results in dense
liquid states are sometimes astonishingly good, even for network
forming liquids\cite{water,SIO2}.

However, the presence of the bare interactions in our formulation of
the theory constitutes a strong limitation to our approach, and, to
our present knowledge, the gaussian approximation, both for static and
dynamics, appears to be the only immediate fully consistent
closure. Thus, the main problem in developing further this formalism
to better types of approximations is how to renormalise more
effectively the interactions, othen than simply RPA. We will attempt
to do so in future work.


\end{document}